\def\year{2020}\relax
\documentclass[letterpaper]{article} 
\usepackage{aaai20}  
\usepackage{times}  
\usepackage{helvet} 
\usepackage{courier}  
\usepackage[hyphens]{url}  
\usepackage{graphicx} 
\usepackage{graphicx}  
\usepackage{booktabs} 
\usepackage{enumitem}
\usepackage{multirow}
\usepackage{cleveref}

\usepackage[dvipsnames]{xcolor}
\usepackage{soul}
\definecolor{mypink1}{rgb}{0.858, 0.188, 0.478}
\definecolor{navy}{rgb}{0.1, 0.1, 0.8}
\definecolor{ruby}{rgb}{0.6, 0.1, 0.3}
\definecolor{ao}{rgb}{0.0, 0.5, 0.0}

\usepackage[colorinlistoftodos,prependcaption,textsize=tiny]{todonotes}

\usepackage{xspace} 
\newcommand{\header}[1]{{\noindent{\textbf{#1}}}}

\newcommand{\cb}{\textsc{Cyberbullying}\xspace}
\newcommand{\yt}{\textsc{YouTube}\xspace}

\newcommand{\citet}[1]{{\citeauthor{#1}~\shortcite{#1}}}

\usepackage{array} 
\usepackage[]{algorithm2e}
\usepackage{multibib}
\newcites{AP}{Appendix References}

\title{Variation across Scales: Measurement Fidelity under Twitter Data Sampling}
\author{Siqi Wu\textsuperscript{\rm 1,3} \and Marian-Andrei Rizoiu\textsuperscript{\rm 2,3} \and Lexing Xie\textsuperscript{\rm 1,3} \\
\textsuperscript{\rm 1}Australian National University, \textsuperscript{\rm 2}University of Technology Sydney, \textsuperscript{\rm 3}Data 61, CSIRO, Australia\\
{\{siqi.wu, lexing.xie\}}@anu.edu.au, marian-andrei.rizoiu@uts.edu.au\\
}

\begin{document}

\maketitle


\begin{abstract}
A comprehensive understanding of data quality is the cornerstone of measurement studies in social media research.
This paper presents in-depth measurements on the effects of Twitter data sampling across different timescales and different subjects (entities, networks, and cascades).
By constructing complete tweet streams, we show that Twitter rate limit message is an accurate indicator for the volume of missing tweets.
Sampling also differs significantly across timescales.
While the hourly sampling rate is influenced by the diurnal rhythm in different time zones, the millisecond level sampling is heavily affected by the implementation choices.
For Twitter entities such as users, we find the Bernoulli process with a uniform rate approximates the empirical distributions well.
It also allows us to estimate the true ranking with the observed sample data.
For networks on Twitter, their structures are altered significantly and some components are more likely to be preserved.
For retweet cascades, we observe changes in distributions of tweet inter-arrival time and user influence, which will affect models that rely on these features.
This work calls attention to noises and potential biases in social data, and provides a few tools to measure Twitter sampling effects.
\end{abstract}




\section{Introduction}
\label{sec:intro}

\begin{quote}
\textit{``Polls are just a collection of statistics that reflect what people are thinking in `reality'. And reality has a well-known liberal bias.''}
-- Stephen Colbert\footnote{At the 2006 White House Correspondents' Dinner.} 
\end{quote}

Data quality is a timely topic that receives broad attention.
The data noises and biases particularly affect data-driven studies in social media~\cite{tufekci2014big,olteanu2019social}.
Overrepresented or underrepresented data may mislead researchers to spurious claims~\cite{ruths2014social}.
For example, opinion polls wrongly predicted the U.S. presidential election results in 1936 and 1948 because of unrepresentative samples~\cite{mosteller1949pre}.
In the era of machine learning, the data biases can be amplified by the subsequent models.
For example, models overly classify agents doing cooking activity as female due to overrepresented correlations~\cite{zhao2017men}, or lack the capacity to identify dark-skinned women due to underrepresented data~\cite{buolamwini2018gender}.
Hence, researchers must be aware and take account of the hidden biases in their datasets for drawing rigorous scientific conclusions.

Twitter is the most prominent data source in ICWSM -- 82 (31\%) out of 265 full papers in the past 5 years (2015-2019) used Twitter data (listed in Section A of \cite{appendix}),
in part because Twitter has relatively open data policies, and in part because Twitter offers a range of public application programming interfaces (APIs).
Researchers have used Twitter data as a lens to understand political elections~\cite{bovet2019influence}, social movements~\cite{de2016social}, information diffusion~\cite{zhao2015seismic}, and many other social phenomena.
Twitter offers two streaming APIs for free, namely \textit{sampled} stream and \textit{filtered} stream.
The filtered stream tracks a set of keywords, users, languages, and locations.
When the matched tweet volume is above a threshold, Twitter subsamples the stream, which compromises the completeness of the collected data.
In this paper, we focus on empirically quantifying the data noises resulted from the sampling in the filtered stream and its impacts on common measurements.

This work addresses two open questions related to Twitter data sampling.
Firstly, \textbf{how are the tweets missing in the filtered stream?}
The sampling mechanism of the sampled stream has been extensively investigated~\cite{kergl2014endogenesis,pfeffer2018tampering}, but relatively little is said about the filtered stream.
Since the two streaming APIs are designed to be used in different scenarios, it is pivotal for researchers who use the filtered stream to understand what, when, and how much data is missing.
Secondly, \textbf{what are the sampling effects on common measurements?}
Our work is inspired by~\citet{morstatter2013sample}, who measured the discrepancies of topical, network, and geographic metrics.
We extend the measurements to entity frequency, entity ranking, bipartite graph, retweet network, and retweet cascades.
The answers to these questions not only help researchers shape appropriate questions, but also help platforms improve their data services.

We address the first question by curating two datasets that track suggested keywords in previous studies.
Without leveraging the costly Twitter Firehose service, we construct the complete tweet streams by splitting the keywords and languages into multiple subcrawlers.
We study the Twitter rate limit messages.
Contradicting observations made by \citet{sampson2015surpassing}, our results show that the rate limit messages closely approximate the volume of missing data.
We also find that tweets are not missing at random since the sampling rates have distinct temporal variations across different timescales, especially at the level of hour and millisecond.

Addressing the second question, we measure the effects of Twitter data sampling across different subjects, e.g., the entity frequency, entity ranking, user-hashtag bipartite graph, retweet network, and retweet cascades.
We find that 
(1) the Bernoulli process with a uniform rate can approximate the empirical entity distribution well;
(2) the ranks of top entities are distorted;
(3) the true entity frequency and ranking can be inferred based on sampled observations;
(4) the network structures change significantly with some components more likely to be preserved; 
(5) sampling compromises the quality of diffusion models as the distributions of tweet inter-arrival time and user influence are substantially skewed.
We remark that this work only studies the effects of Twitter sampling mechanism, but does not intend to reverse engineer it.

The main contributions of this work include:
\begin{itemize}[leftmargin=*]
  \item We show that Twitter rate limit message is an accurate indicator for the volume of missing tweets.
  \item A set of measurements on the Twitter data sampling effects across different timescales and different subjects.
  \item We show how to estimate the entity frequency and ranking of the complete data using only the sample data.
  \item We release a software package ``Twitter-intact-stream'' for constructing the complete data streams on Twitter\footnote{The package, collected data, and analysis code are publicly available at \url{https://github.com/avalanchesiqi/twitter-sampling}}.
\end{itemize}




\section{Related work}
\label{sec:related}

\header{Studies on Twitter APIs.}
Twitter has different levels of access (Firehose, Gardenhose, Spritzer) and different ways to access (search API, sampled stream, filtered stream). 
As the complete data service (Firehose) incurs excessive costs and requires severe storage loads, we only discuss the free APIs.

\begin{itemize}[leftmargin=*]
  \item Twitter search API returns relevant tweets for a given query, but it only fetches results published in the past 7 days~\cite{twittersearch}.
  The search API also bears the issue of data attrition.
  Research using this API to construct a ``complete'' dataset would inevitably miss parts of desired tweets~\cite{wang2015should} since tweet creation and deletion are highly dynamic~\cite{almuhimedi2013tweets}.
  To overcome this limitation, researchers can pivot to the streaming APIs.
  
  \item Twitter sampled streaming API returns roughly 1\% of all public tweets in realtime~\cite{twittersample}.
  \citet{pfeffer2018tampering} detailed its sampling mechanism and identified potential tampering behaviors.
  \citet{gonzalez2014assessing} examined the biases in the retweet network from the 1\% sample and the search API.
  While the 1\% sample may be treated as a representative sample of overall Twitter activities~\cite{morstatter2014biased,kergl2014endogenesis}, data filtering can only be conducted post data collection.
  Therefore, it is not suitable to create ad hoc datasets, e.g., tracking \textit{all} tweets that contain the hashtag \texttt{\#metoo}.
  
  \item Twitter filtered streaming API collects tweets matching a set of prescribed predicates in realtime~\cite{twitterfilter}.
  Suppose that the streaming rate is below Twitter limit, the pre-filtering makes the filtered stream possible to construct the complete datasets without using the costly Firehose stream, e.g., on social movements~\cite{de2016social} and on news outlets~\cite{mishra2016feature}.
  We focus on the scenes where the data streams are sampled.
  The most relevant work is done by \citet{morstatter2013sample}, in which they compared the filtered stream to the Firehose, and measured the discrepancies in various metrics.
  We extend the scope of measured subjects.
  Furthermore, we take a step to correct the sampling effects on entity measures.
\end{itemize}

Twitter sampling is deterministic~\cite{joseph2014two}, therefore, simply stacking crawlers with the same predicates will not yield more data.
However, users can improve the sample coverage by splitting the keyword set into multiple disjoint predicate sets, and monitoring each set with a distinct subcrawler~\cite{sampson2015surpassing}.

\header{Effects of missing social data.}
Social data, which records ubiquitous human activities in digital form, plays a fundamental role in social media research.
Researchers have pointed out the necessity to interrogate the assumptions and biases in data~\cite{boyd2012critical,ruths2014social}.
\citet{tufekci2014big} outlined four issues on data representativeness and validity.
The hidden data biases may alter some research conclusions and even impact human decision making~\cite{olteanu2019social}.

\citet{gaffney2018caveat} identified gaps where data is unevenly missing in a widely used Reddit corpus.
They suggested strong risks in research that concerns user history or network information, and moderate risks in research that uses aggregate counts.
In this work, we use these qualitative observations as starting points and present a set of in-depth quantitative measurements.
We corroborate the risks in user history study and network analysis.
And we show how the complete counting statistics can be estimated.

\header{Sampling from graphs and cascades.}
\citet{leskovec2006sampling} studied different graph sampling strategies for drawing representative samples.
\citet{wagner2017sampling} considered how sampling impacts the relative ranking of groups in the attributed graphs.
The effects of graph sampling has been extensively discussed by \citet{kossinets2006effects}.
In this work, the missing tweets can cause edge weights to decrease, and some edges to even disappear.
On sampling a cascade, \citet{de2010does} found that combining network topology and contextual attributes distorts less the observed metrics.
\citet{sadikov2011correcting} proposed a \textit{k}-tree model to uncover some properties from the sampled data.
They both sampled the cascades via different techniques (e.g., random, forest fire) and varying ratios.
In contrast, the sampling in this work is an artifact of proprietary Twitter sampling mechanisms, and beyond the control of the users.




\begin{figure*}[tbp]
    \centering
	\includegraphics[width=2.09\columnwidth]{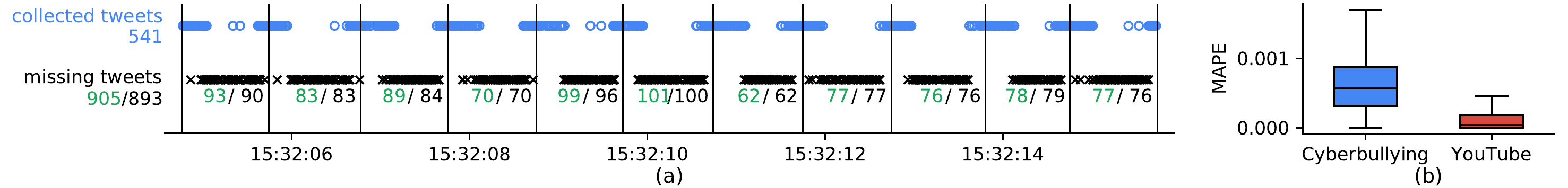}
	\caption{\textbf{(a)} Collected and missing tweets in an 11-second interval. \textcolor{blue}{blue} circle: collected tweet; black cross: missing tweet; black vertical line: rate limit message. \textcolor{ForestGreen}{green} number: estimated missing volume from rate limit messages; black number: count of missing tweets compared to the complete set.
	\textbf{(b)} MAPE of estimating the missing volumes in the rate limit segments.}
	\label{fig:data-validate-ratemsg}
\end{figure*}

\section{Datasets and Twitter rate limit messages}
\label{sec:data}

We collect two datasets, using two sets of keywords employed in recent large-scale studies that use Twitter.
We choose these works because they are high volume and informative for important social science problems (cyberbullying and online content sharing).
We use $\rho$ to denote the sampling rate -- i.e., the probability that a tweet is present in the collected (sampled) dataset.
We use subscripts to differentiate sampling rates that vary over time $\rho_t$, users $\rho_u$, networks $\rho_n$, and cascades $\rho_c$.
The datasets are collected using the Twitter filtered streaming API and are summarized in \Cref{table:dataset}.

\begin{itemize}[leftmargin=*]
  \item \cb~\cite{nand2016bullying}: This dataset tracks all tweets that mention any of the 25 recommended keywords from psychology literature (e.g., \texttt{gay}, \texttt{slut}, \texttt{loser}). The collection period is from 2019-10-13 to 2019-10-26.
  \item \yt~\cite{rizoiu2017expecting}: This dataset tracks all tweets that contain at least one YouTube video URL by using the rule ``\texttt{youtube}'' \texttt{OR} (``\texttt{youtu}'' \texttt{AND} ``\texttt{be}''). The collection period is from 2019-11-06 to 2019-11-19.
\end{itemize}

The streaming client is a program that receives streaming data via Twitter API.
The client will be rate limited if the number of matching tweets exceeds a preset threshold -- 50 tweets per second as of 2020-03~\cite{twitterrate2}.
When we use only one client to track all keywords, we find that both datasets trigger rate limiting.
We refer to the crawling results from a single client as the \textit{sample set}.

We develop a software package ``Twitter-intact-stream''\footnote{\url{https://github.com/avalanchesiqi/twitter-intact-stream}} for constructing the complete data streams on Twitter.
The package splits the filtering predicates into multiple subsets, and tracks each set with a distinct streaming client.
The \cb and \yt datasets are respectively crawled by 8 and 12 clients based on different combinations of keywords and languages.
We remove the duplicate tweets and sort the distinct tweets chronologically.
We refer to the crawling results from multiple clients as the \textit{complete set}.

In very occasional cases, the complete sets also encounter rate limiting.
Estimated from the rate limit messages (detailed next), 0.04\% and 0.14\% tweets in the complete sets are missing, which are negligible comparing to the volumes of missing tweets in the sample sets (47.28\% and 8.47\%, respectively).
For rigorous comparison, we obtain a 30 minutes complete sample from Twitter Firehose and find the difference with our collected data is trivial (detailed in Section C of \cite{appendix}).
Hence, for the rest of this work, we treat the complete sets as if they contain no missing tweets.



\begin{table}
    \centering
    \resizebox{.99\columnwidth}{!}{
    \begin{tabular}{rrrrr}
        \toprule
        & \multicolumn{2}{c}{\cb} & \multicolumn{2}{c}{\yt} \\
        & complete & sample & complete & sample \\
        \midrule
        $N_c$ & 114,488,537 & 60,400,257 & 53,557,950 & 49,087,406 \\
        $N_r$ & 3,047 & 1,201,315 & 3,061 & 320,751 \\
        $\hat{N}_m$ & 42,623 & 54,175,503 & 77,055 & 4,542,397 \\
        $\bar{\rho}$ & 99.96\% & 52.72\% & 99.86\% & 91.53\% \\
        \bottomrule
    \end{tabular}
    }
    \caption{Summary of two datasets. $N_c$: \#collected tweets; $N_r$: \#rate limit messages; $\hat{N}_m$: \#estimated missing tweets; $\bar{\rho}$: mean sampling rate. Full specifications for all streaming clients are listed in Section B of \cite{appendix}.}
    \label{table:dataset}
\end{table}


\header{Validating Twitter rate limit messages.}
When the streaming rate exceeds the threshold, Twitter API emits a rate limit message that consists of a timestamp and an integer.
The integer is designed to indicate the cumulative number of missing tweets since the connection starts~\cite{twitterrate1}.
Therefore, the difference between 2 consecutive rate limit messages should estimate the missing volume in between.

We empirically validate the rate limit messages.
We divide the datasets into a list of segments where (a) they contain no rate limit message in the complete set; (b) they are bounded by 2 rate limit messages in the sample set.
This yields 1,871 and 253 segments in the \cb and \yt datasets, respectively.
The lengths of segments range from a few seconds to several hours, and collectively cover 13.5 days out of the 14-day crawling windows.
In this way, we assure that the segments in the complete set have no tweet missing since no rate limit message is received.
Consequently, for each segment we can compute the volume of missing tweets in the sample set by either computing the difference of the two rate limit messages bordering the segment, or by comparing the collected tweets with the complete set.
\Cref{fig:data-validate-ratemsg}(a) illustrates the collected and missing tweets in an 11-second interval.
The estimated missing volumes from rate limit messages closely match the counts of the missing tweets in the complete set.
Overall, the median error in estimating the missing volume using rate limit messages is less than 0.0005, measured by mean absolute percentage error (MAPE) and shown in \Cref{fig:data-validate-ratemsg}(b).
We thus conclude that the rate limit message is an accurate indicator for the number of missing tweets.
Note that it only approximates the volume of missing tweets, but not the content.

Our observations contradict those from \citet{sampson2015surpassing}, who used the same keyword-splitting approach, yet found that the rate limit messages give inaccurate estimations.
They consistently retrieved more distinct tweets (up to 2 times) than the estimated total volume, i.e., the number of collected tweets plus the estimated missing tweets.
In contrast, our datasets only have a small deviation (0.08\% and 0.13\%, comparing $N_c$ of the complete set to $N_c{+}\hat{N}_m$ of the sample set in \Cref{table:dataset}).
This discrepancy is due to a different implementation choice back in 2015 -- instead of having 1 rate limit message for each second, the rate limit messages were spawned across 4 threads, resulting in up to 4 messages per second.
We provide a detailed analysis in Section C of \cite{appendix}.



\section{Are tweets missing at random?}
\label{sec:random}


In this section, we study the randomness of Twitter sampling
-- do all tweets share the same probability of missing?
This is relevant because uniform random sampling creates representative samples.
When the sampling is not uniform, the sampled set may suffer from systematic biases, e.g., some tweets have a higher chance of being observed.
Consequently, some users or hashtags may appear more often than their cohorts.
We tackle the uniformity of the sampling when accounting for the tweet timestamp, language, and type.


\begin{figure}[tbp]
    \centering
    \includegraphics[width=.99\columnwidth]{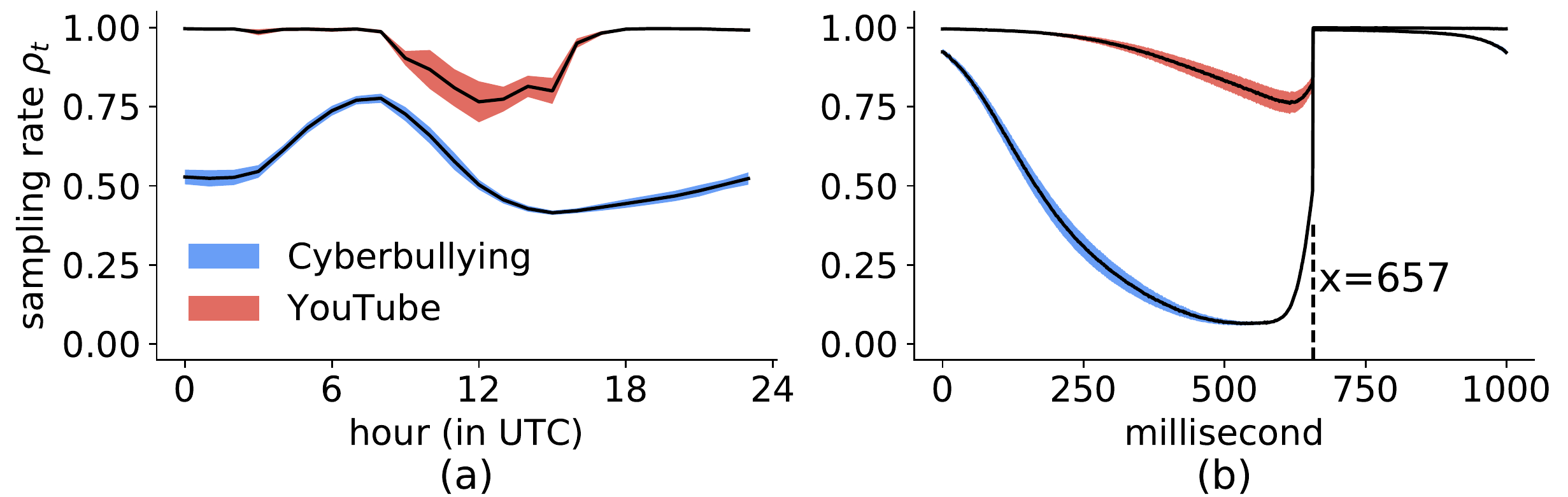} \\
	\caption{Sampling rates are uneven \textbf{(a)} in different hours or \textbf{(b)} in different milliseconds.
	black line: temporal mean sampling rates; color shades: 95\% confidence interval.}
	\label{fig:temporal-sampling-rates}
\end{figure}

\header{Tweet timestamps.}
\Cref{fig:temporal-sampling-rates}(a) plots the hourly sampling rates.
\cb dataset has the highest sampling rate ($\rho_t{=}78\%$) at UTC-8.
The lowest sampling rate ($\rho_t{=}41\%$) occurs at UTC-15, about half of the highest value.
\yt dataset is almost complete ($\rho_t{=}100\%$) apart from UTC-8 to UTC-17.
The lowest sampling rate is $76\%$ at UTC-12.
We posit that the hourly variation is related to the overall tweeting dynamics and the rate limit threshold (i.e., 50 tweets per second):
higher tweet volumes yield lower sampling rates.
\Cref{fig:temporal-sampling-rates}(b) shows the sampling rate at the millisecond level, which curiously exhibits a periodicity of one second.
In \cb dataset, the sampling rate peaks at millisecond 657 ($\rho_t{=}100\%$) and drops monotonically till millisecond 550 ($\rho_t{=}6\%$) before bouncing back.
\yt dataset follows a similar trend with the lowest value ($\rho_t{=}76\%$) at millisecond 615.
This artifact leaves the sample set vulnerable to automation tools.
Users can deliberately schedule tweet posting time within the high sampling rate period for inflating their representativeness, or within the low sampling rate period for masking their content in the public API.
The minutely and secondly sampling rates are included in Section D of \cite{appendix}.


\begin{figure}[tbp]
    \centering
    \includegraphics[width=.99\columnwidth]{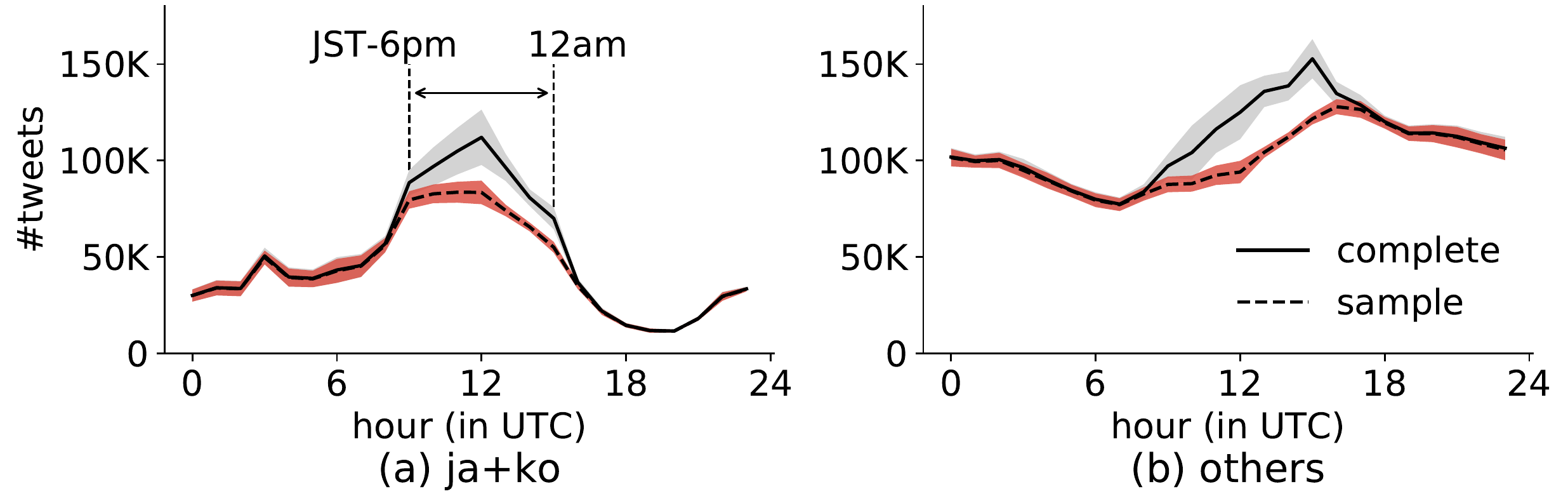} \\ 
	\caption {Hourly tweet volumes in \yt dataset. \textbf{(a)} Japanese+Korean; \textbf{(b)} other languages.
	black line: temporal mean tweet volumes; color shades: 95\% confidence interval.}
	\label{fig:tweet-lang-vol}
\end{figure}


\begin{table}
    \centering
    \resizebox{.99\columnwidth}{!}{
    \begin{tabular}{rrrrr}
        \toprule
        & \multicolumn{2}{c}{\cb} & \multicolumn{2}{c}{\yt} \\
        & complete & sample & complete & sample  \\
        \midrule
        \%root tweets & 14.28\% & 14.26\% & 25.90\% & 26.19\% \\
        \%retweets & 64.40\% & 64.80\% & 62.92\% & 62.51\% \\
        \%quotes & 7.37\% & 7.18\% & 3.44\% & 3.40\% \\
        \%replies & 13.94\% & 13.76\% & 7.74\% & 7.90\% \\
        \bottomrule
    \end{tabular}
    }
    \caption{The ratios of the 4 tweet types (root tweet, retweet, quote, and reply) in the complete and the sample sets.}
    \label{table:type}
\end{table}

\header{Tweet languages.}
Some languages are mostly used within one particular timezone, e.g., Japanese and Korean\footnote{Japanese Standard Time (JST) and Korean Standard Time (KST) are the same.}.
The temporal tweet volumes for these languages are related to the daily activity patterns in the corresponding countries.
We break down the hourly tweet volumes of \yt dataset into Japanese+Korean and other languages.
The results are shown in \Cref{fig:tweet-lang-vol}.
Altogether, Japanese and Korean account for 31.4\% tweets mentioning YouTube URLs.
The temporal variations are visually different -- 48.3\% of Japanese and Korean tweets are posted in the evening of local time (JST-6pm to 12am), while tweets in other languages disperse more evenly.
Because of the high volume of tweets in this period,  sampling rates within UTC-9 to UTC-15 are lower (see \Cref{fig:temporal-sampling-rates}(a)).
Consequently, ``ja+ko'' tweets are less likely to be observed (89.0\% in average, 80.9\% between JST-6pm and 12am) than others (92.9\% in average).

\header{Tweet types.}
Twitter allows the creation of 4 types of tweets.
The users create a \textit{root tweet} when they post new content from their home timelines.
The other 3 types are interactions with existing tweets: \textit{retweets} (when users click on the ``Retweet'' button); \textit{quotes} (when users click on the ``Retweet with comment'' button); \textit{replies} (when users click on the ``Reply'' button).
The relative ratios of different types of tweets are distinct for the two datasets (see \Cref{table:type}).
\cb has higher ratios of retweets, quotes, and replies than \yt, implying more interactions among users.
However, the ratios of different types are very similar in the sampled versions of both datasets (max deviation$=$0.41\%, retweets in \yt dataset).
We conclude that Twitter data sampling is not biased towards any tweet type.



\section{Impacts on Twitter entities}
\label{sec:entity}

In this section, we study how the data sampling affects the observed frequency and relative ranking of Twitter entities, e.g., users, hashtags, and URLs.
We first use a Bernoulli process to model the Twitter data sampling (\Cref{ssec:bernoulli}).
Next, we show how the entity statistics for one set (e.g., the complete) can be estimated using the other set (the sample, \Cref{ssec:frequency}).
Finally, we measure the distortions introduced in entity ranking by sampling and how to correct them (\Cref{ssec:ranking}).
The basic statistics of entities are listed in~\Cref{table:entity}.
The analyses in this section, \Cref{sec:network}, and \Cref{sec:cascade}, are done with \cb dataset since its sampling effects are more prominent.


\begin{table}
    \centering
    \resizebox{.99\columnwidth}{!}{
    \begin{tabular}{rrrrr}
        \toprule
        & complete & sample & \%miss & est. \%miss \\
        \midrule
        \#users & 19,802,506 & 14,649,558 & 26.02\% & 26.12\% \\
        \#hashtags & 1,166,483 & 880,096 & 24.55\% & 24.31\% \\
        \#URLs & 467,941 & 283,729 & 39.37\% & 38.99\% \\
        \bottomrule
    \end{tabular}
    }
    \caption{Statistics of entities in \cb dataset, mean sampling rate $\bar{\rho}{=}52.72\%$.}
    \label{table:entity}
\end{table}

\subsection{Twitter sampling as a Bernoulli process}
\label{ssec:bernoulli}


\begin{figure}[tbp]
    \centering
    \includegraphics[width=.99\columnwidth]{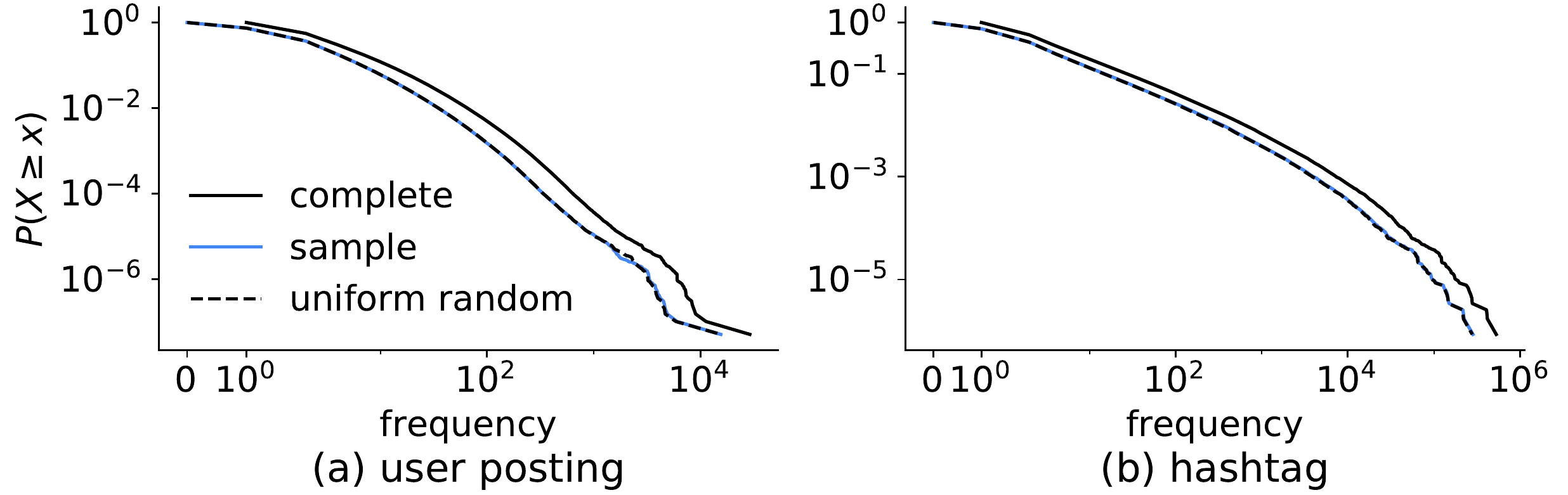}
	\caption{The frequency distributions of \textbf{(a)} user posting and \textbf{(b)} hashtag. The x-axis starts at 0 rather than 1, as the sample set and uniform random sample both have missing entities.}
	\label{fig:entity-frequency}
\end{figure}

We examine how well we can use a Bernoulli process to approximate the Twitter sampling process.
Assuming that tweets are sampled identically and independently, the Twitter sampling can be be seen as a simple Bernoulli process with the mean sampling rate $\bar{\rho}$.
We empirically validate this assumption by plotting the complementary cumulative density functions (CCDFs) of user posting frequency (the number of times a user posts) and hashtag frequency (the number of times a hashtag appears) in \Cref{fig:entity-frequency}.
The black and \textcolor{blue}{blue} solid lines respectively show the CCDFs of the complete and the sample sets, while the black dashed line shows the CCDF in a synthetic dataset constructed from the complete set using a Bernoulli process with rate $\bar{\rho}{=}52.72\%$.
Firstly, we observe that the CCDF of the sample set is shifted left, towards the lower frequency end.
Visually, the distributions for the synthetic (black dashed line) and for the observed sample set (\textcolor{blue}{blue} solid line) overlap each other.
Furthermore, following the practices in~\cite{leskovec2006sampling}, we measure the agreement between these distributions with Kolmogorov-Smirnov D-statistic, which is defined as
\begin{equation}
	D(G, G^\prime) = \mathrm{max}_x \{|G(x) - G^\prime(x)|\}
\end{equation}
where $G$ and $G^\prime$ are the cumulative distribution functions (CDFs) of two distributions.
With a value between 0 and 1, a smaller D-statistic implies more agreement between two measured distributions.
The results show high agreement between entity distributions in the synthetic and the observed sample sets (0.0006 for user posting and 0.002 for hashtag).
This suggests that despite the empirical sampling rates not being unique over time, a Bernoulli process of constant rate can model the observed entity frequency distribution well\footnote{We do not choose the goodness of fit test (e.g., Kolmogorov-Smirnov test) because our sample sizes are in the order of millions. And trivial effects can be found to be significant  with very large sample sizes. Instead we report the effect sizes (e.g., D-statistic). Alternative distance metrics (e.g., Bhattacharyya distance or Hellinger distance) yield qualitatively similar results.}.

\subsection{Entity frequency}
\label{ssec:frequency}

We investigate whether the statistics on one set (complete or sample) can be estimated using only the statistics of the other set and the Bernoulli process model.
We use $n_c$ to denote the frequency in the complete set, and $n_s$ the frequency in the sample set ($n_c{\geq}n_s$).
More precisely, we ask these three questions:
What is the distribution of $n_s$ given $n_c{=}k$?
What is the distribution of $n_c$ given $n_s{=}k$?
How many entities are missing altogether given the distribution of $n_s$?






\begin{figure}[tbp]
    \centering
    \includegraphics[width=.99\columnwidth]{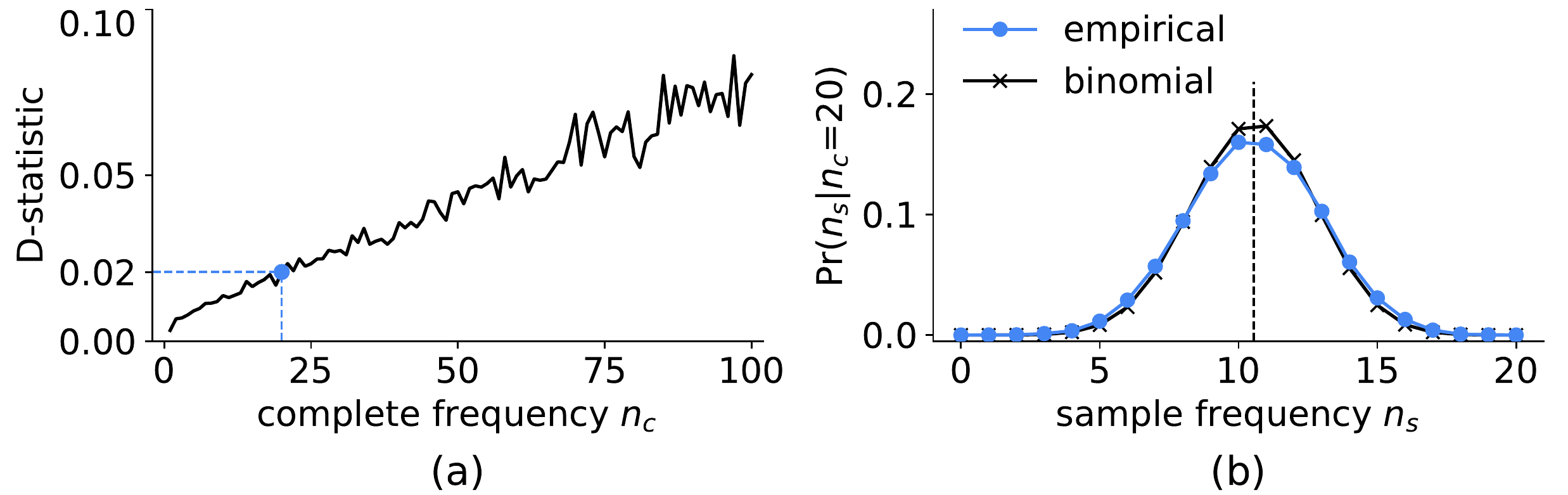}
	\caption{\textbf{(a)} D-statistic between empirical distribution and binomial distribution for the number of tweets a user posts.
 	\textbf{(b)} The probability distribution of observing $n_s$ times in the sample set when $n_c{=}20$.}
	\label{fig:measure-entity-binomial}
\end{figure}

\header{Modeling sample frequency from the complete set.}

For a user who posts $n_c$ times in the complete set, their sample frequency under the Bernoulli process follows a binomial distribution $B(n_c, \bar{\rho})$.
Specifically, the probability of observing the user $n_s$ times in the sample set is 
\begin{equation}
\mathrm{Pr}(n_s | n_c, \bar{\rho}) = {n_c \choose n_s} \bar{\rho}^{n_s} (1{-}\bar{\rho})^{n_c - n_s}
\end{equation}
We compute the empirical distribution and binomial distribution for $n_c$ from 1 to 100.
This covers more than 99\% users in our dataset.
\Cref{fig:measure-entity-binomial}(a) shows the D-statistic between two distributions as a function of complete frequency $n_c$.
The binomial distribution models the empirical data better when $n_c$ is smaller.
\Cref{fig:measure-entity-binomial}(b) illustrates an example of $n_c{=}20$.
The binomial distribution closely approximates the empirical distribution.
Their mean sample frequencies (dashed vertical lines) are also identical (10.54).





\header{Inferring complete frequency from the sample set.}
Under the Bernoulli process, for users who are observed $n_s$ times in the sample set, their complete frequencies follows a negative binomial distribution $NB(n_s, \bar{\rho})$.
The negative binomial distribution models the discrete probability distribution of the number of Bernoulli trials before a predefined number of successes occurs.
In our context, given $n_s$ tweets ($n_s{\geq}1$) are successfully sampled, the probability of having $n_c$ tweets in the complete set is
\begin{equation}
\mathrm{Pr}(n_c | n_s, \bar{\rho}) = {n_c{-}1 \choose n_s{-}1} \bar{\rho}^{n_s} (1{-}\bar{\rho})^{n_c - n_s}
\end{equation}
%


\begin{figure}[tbp]
    \centering
    \includegraphics[width=.99\columnwidth]{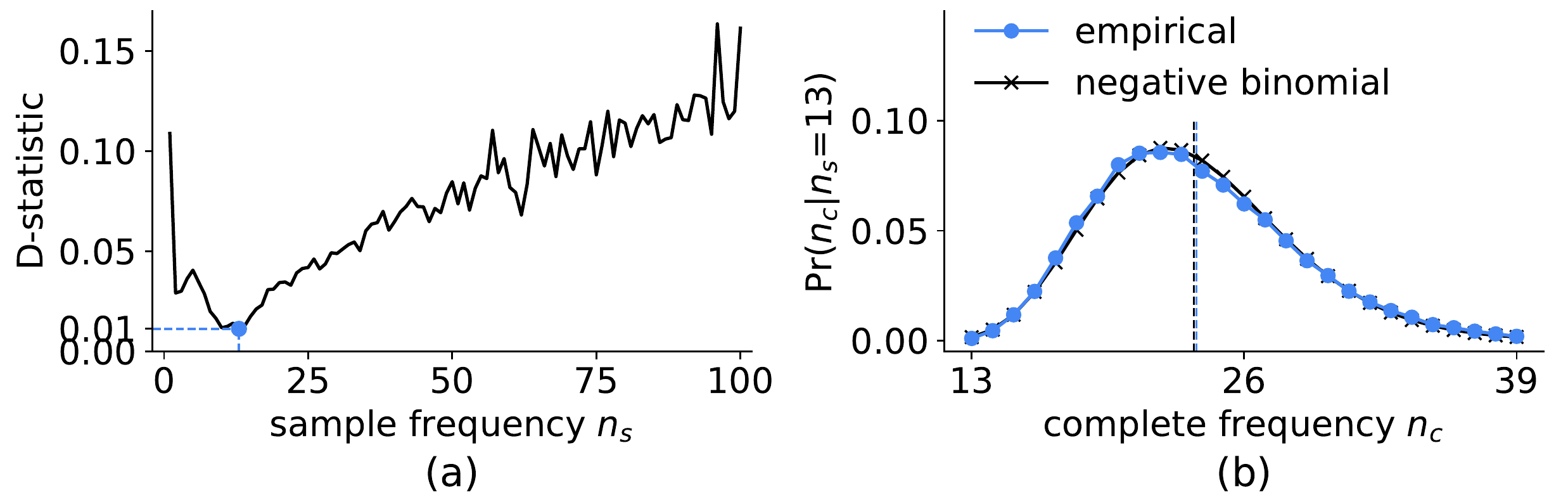}
	\caption{\textbf{(a)} D-statistic between empirical distribution and negative binomial distribution for the number of tweets a user posts.
	\textbf{(b)} The probability distribution of posting $n_c$ times in the complete set when $n_s{=}13$.}
    \label{fig:entity-negbin}
\end{figure}

We compute the empirical distribution and negative binomial distribution for $n_s$ from 1 to 100.
\Cref{fig:entity-negbin}(a) shows the D-statistic as a function of sample frequency $n_s$.
Negative binomial distributions models the best when the number of observed tweets is between 9 and 15 (D-statistic${<}$0.02).
\Cref{fig:entity-negbin}(b) shows both distributions for $n_s{=}13$, where the minimal D-statistic is reached.
The negative binomial distribution closely resembles the empirical distribution.
Their estimated mean complete frequencies are very similar (23.60 vs. 23.72, shown as dashed vertical lines).


\header{Estimating missing volume from the sample set.}
In data collection pipelines, the obtained entities from the filtered stream are sometimes used as seeds for the second step crawling, such as constructing user timelines based on user IDs~\cite{wang2015should}, or querying YouTube statistics based on video URLs~\cite{wu2018beyond}.
However, some entities may be completely missing due to Twitter sampling.
We thus ask: can we estimate the total number of missing entities given the entity frequency distribution of the sample set?

We formulate the problem as solving a matrix equation with constraints.
We use the symbol $\textbf{F}$ to denote the entity frequency vector.
$\textbf{F}[n_s]$ represents the number of entities that occurs $n_s$ times in the sample set.
We want to estimate the frequency vector $\hat{\textbf{F}}$ of the complete set.
For any $n_s$, its sample frequency $\textbf{F}[n_s]$ satisfies
\begin{equation}
\textbf{F}[n_s] = \sum_{k{=}n_s}^{\infty}\mathrm{Pr}(n_s | k, \bar{\rho}) * \hat{\textbf{F}}[k]
\end{equation}
We constrain $\hat{\textbf{F}}$ to be non-negative numbers and decrease monotonically since the frequency distribution is usually heavy-tailed in practice (see~\Cref{fig:entity-frequency}).
We use the frequency vector for $n_s{\in}[1, 100]$.
The above matrix equation can be solved as a constrained optimization task.
For users who post $n_c$ times in the complete set, the probability of their tweets completely missing is $\mathrm{Pr}(n_s{=}0; n_c, \bar{\rho}){=}(1{-}\bar{\rho})^{n_c}$.
Altogether, the estimated missing volume is $\sum_{n_c=1}^{\infty}{(1{-}\bar{\rho})^{n_c} \hat{\textbf{F}}[n_c]}$ for the whole dataset.
We show the estimated results in the rightmost column of \Cref{table:entity}.
The relative errors (MAPE) are smaller than 0.5\% for all entities.
This suggests that the volume of missing entities can be accurately estimated if the frequency distribution of the sample set is observed.

\header{Summary.}
Although the empirical sample rates have clear temporal variations, we show that we can use the mean sampling rate to estimate some entity statistics, including the frequency distribution and the missing volume.
This reduces the concerns on assuming the observed data stream is a result of uniform random sampling~\cite{joseph2014two,morstatter2014biased,pfeffer2018tampering}.

\subsection{Entity ranking}
\label{ssec:ranking}
Entity ranking is important for many social media studies.
One of the most common strategies in data filtering is to keep entities that rank within the top $x$, e.g., most active users or most mentioned hashtags~\cite{morstatter2013sample,gonzalez2014assessing}.
We measure how the Twitter data sampling distorts entity ranking for the most active users, and whether the ground-truth ranking in the complete set can be inferred from the sample ranking.
Note that in this subsection, we allow the sampling rates to be time-dependent $\rho_t$ and user-dependent $\rho_u$ -- as the sampling with a constant rate would preserve the ranking between the complete and the sample sets.
For the universal ranking (considering all entities), we use percentile to measure it and find the higher ranked entities have smaller deviations (detailed in Section E of \cite{appendix}).





\begin{figure}[tbp]
    \centering
    \includegraphics[width=.99\columnwidth]{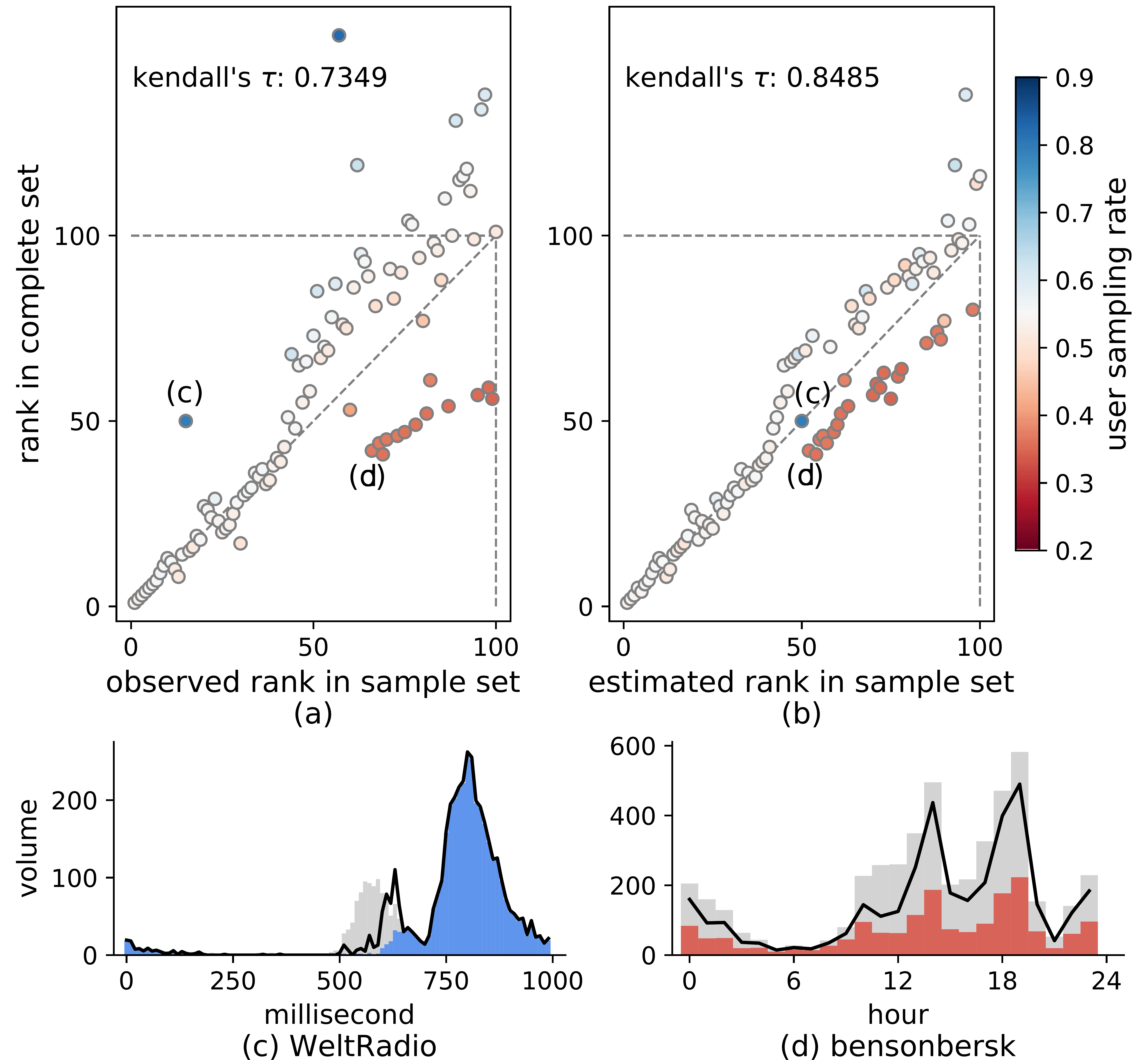}
	\caption{\textbf{(a)} Observed ranks in the sample set (x-axis) vs. true ranks in the complete set (y-axis).
	\textbf{(b)} Estimated ranks improve the agreement with the ground-truth ranks.
	\textbf{(c)} user \textit{WeltRadio}, observed/true/estimated ranks: 15/50/50.
	\textbf{(d)} user \textit{bensonbersk}, observed/true/estimated ranks: 66/42/52.
	\textcolor{blue}{blue}/\textcolor{red}{red} shades: sample tweet volume; grey shades: complete tweet volume; black line: estimated tweet volume.}
	\label{fig:top-rank}
\end{figure}

\header{Detecting rank distortion.}
\Cref{fig:top-rank}(a) plots the most active 100 users in the sample set on the x-axis, and their ranks in the complete set on the y-axis.
Each circle is colored based on the corresponding user sampling rate $\rho_u$.
The diagonal line indicates uniform random sampling, in which the two sets of ranks should be preserved.
The users above the diagonal line improve their ranks in the sample set, while the ones below lose their positions.
\Cref{fig:top-rank}(c) highlights a user \textit{WeltRadio}, who benefits the most from the sampling:
it ranks 50th in the complete set, but it is boosted to 15th place in the sample set.
Comparing the complete tweet volume, its volume (4,529) is only 67\% relative to the user who actually ranks 15th in the complete set (6,728, user \textit{thirdbrainfx}).
We also find that \textit{WeltRadio} tweets mostly in the very high sampling rate secondly period (millisecond 657 to 1,000), resulting in a high user sampling rate ($\rho_u{=}79.1\%$).
On the contrary, \Cref{fig:top-rank}(d) shows a user \textit{bensonbersk} with decreased rank in the sample set and low sampling rate ($\rho_u{=}36.5\%$).
Examining his posting pattern, this user mainly tweets in the low sampling rate hours (UTC-12 to 19).


\header{Estimating true ranking from the sample set.}
Apart from measuring the rank distortion between the complete and the sample sets, we investigate the possibility of estimating the ground-truth ranks by using the observations from the sample set.
From the rate limit messages, we extract the temporal sampling rates that are associated with different timescales (hour, minute, second, and millisecond), i.e., $\rho_t(h,m,s,ms)$.
Based on the negative binomial distribution, for a user who we observe $n_s$ times at timestamp $\kappa{=}(h,m,s,ms)$, the expected volume is $n_s / \rho_t(\kappa)$.
We compute the estimated tweet volumes for all users and select the most active 100 users.
\Cref{fig:top-rank}(b) shows the estimated ranks on the x-axis and the true ranks on the y-axis.
We quantify the degree of agreement using Kendall's $\tau$, which computes the difference of concordant and discordant pairs between two ranked lists.
With value between 0 and 1, a larger value implies more agreement.
The Kendall's $\tau$ is improved from 0.7349 to 0.8485 with our estimated ranks.
The rank correction is important since it allows researchers to mitigate the rank distortion without constructing a complete data stream.



\section{Impacts on networks}
\label{sec:network}


In this section, we measure the effects of data sampling on two commonly studied networks on Twitter: the user-hashtag bipartite graph, and the user-user retweet network.


\subsection{User-hashtag bipartite graph}
\label{ssec:bipartite}

The bipartite graph maps the affiliation between two disjoint sets of entities.
No two entities within the same set are linked.
Bipartite graphs have been used in many social applications, e.g., mining the relation between scholars and published papers~\cite{newman2001structure}, or between artists and concert venues~\cite{arakelyan2018mining}.
Here we construct the user-hashtag bipartite graphs for both the complete and the sample sets.
This graph links users to their used hashtags.
Each edge has a weight -- the number of tweets between its associated user and hashtag.
The basic statistics for the bipartite graphs are summarized in \Cref{table:bipartite}.

\begin{table}
    \centering
    \resizebox{.99\columnwidth}{!}{
    \begin{tabular}{rrrr}
        \toprule
         & complete & sample & ratio \\
        \midrule
        \#tweets with hashtags & 24,539,003 & 13,149,980 & 53.59\% \\
        \#users with hashtags & 6,964,076 & 4,758,161 & 68.32\% \\
        avg. hashtags per user & 9.23 & 7.29 & 78.97\% \\
        \#hashtags & 1,166,483 & 880,096 & 75.45\% \\
        avg. users per hashtags & 55.09 & 39.40 & 71.51\% \\
        \bottomrule
    \end{tabular}
    }
    \caption{Statistics of user-hashtag bipartite graph in \cb dataset. Ratio (rightmost column) compares the value of the sample set against that of the complete set, mean sampling rate $\bar{\rho}{=}52.72\%$.}
    \label{table:bipartite}
\end{table}


\begin{figure}[tbp]
    \centering
    \includegraphics[width=.99\columnwidth]{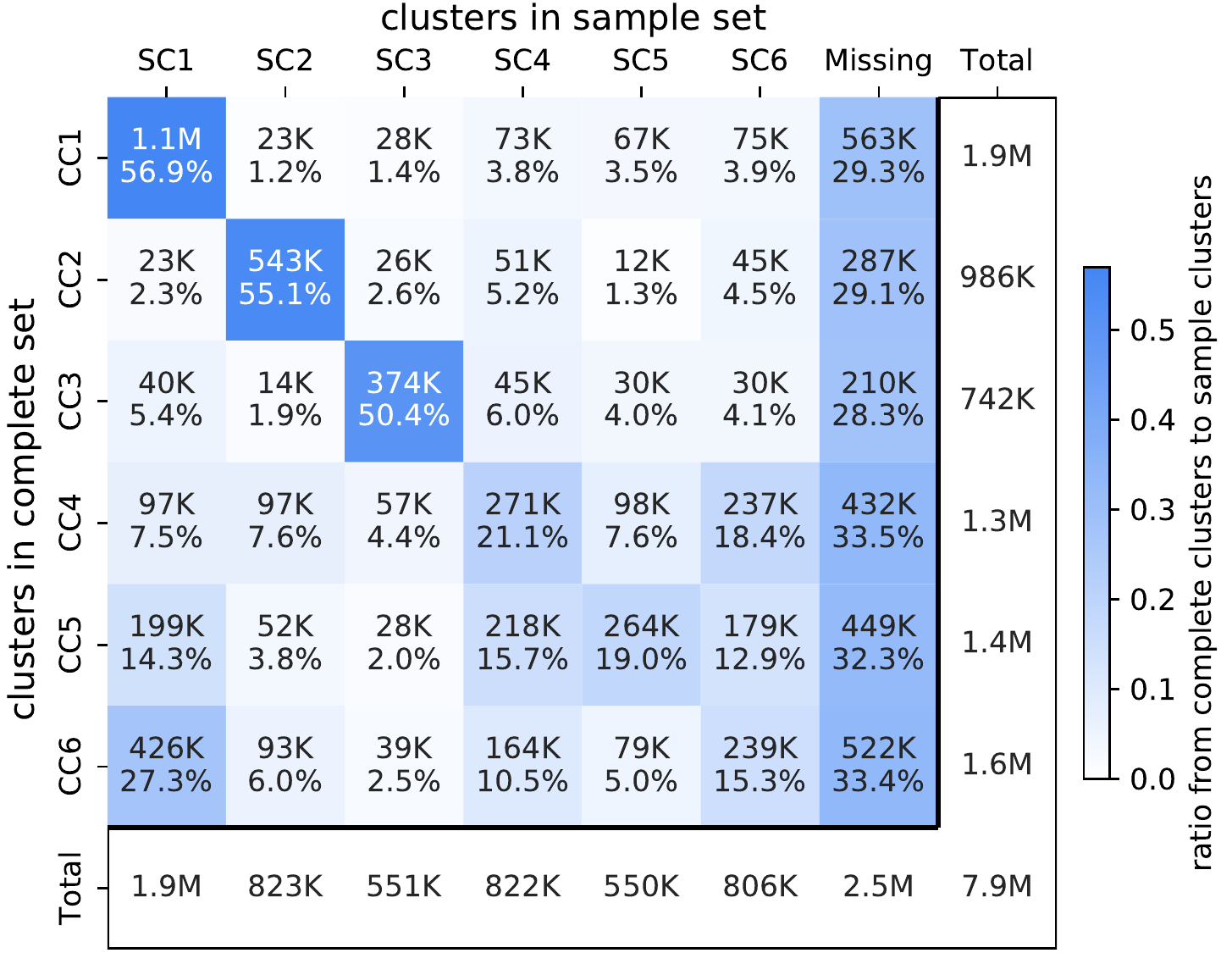}
	\caption{The change of clusters from complete set to sample set.
	Each cell denotes the volume (top number) and the ratio (bottom percentage) of entities (users and hashtags) that traverse from a complete cluster to a sample cluster.
	Clusters are ordered to achieve maximal ratios along the diagonal.}
	\label{fig:bipartite-cluster}
\end{figure}


\begin{table*}
    \centering
    \resizebox{2.09\columnwidth}{!}{
    \begin{tabular}{p{0.3cm}rcccccc}
        \toprule
        \multirow{11}{*}{\rotatebox[origin=c]{90}{complete set}} & & \textbf{CC1} & \textbf{CC2} & \textbf{CC3} & CC4 & CC5 & CC6\\
        & size & \textbf{1,925,520} & \textbf{986,262} & \textbf{742,263} & 1,289,086 & 1,389,829 & 1,562,503  \\
        & \#users & \textbf{1,606,450} & \textbf{939,288} & \textbf{602,845} & 1,080,359 & 1,227,127 & 1,390,276 \\
        & \#hashtags & \textbf{319,070} & \textbf{46,974} & \textbf{139,418} & 208,727 & 162,702 & 172,227 \\
        & avg. degree & \textbf{8.03} & \textbf{7.64} & \textbf{22.19} & 3.46 & 4.74 & 4.07 \\
        & category & \textbf{politics} & \textbf{Korean pop} & \textbf{cyberbullying} & Southeast Asia pop & politics & streaming \\
        \cmidrule{2-8}
        & \multirow{5}{*}{hashtags} & \textbf{brexit} & \textbf{bts} & \textbf{gay} & peckpalitchoke(th) & kamleshtiwari & ps4live \\
        & & \textbf{demdebate} & \textbf{mamavote} & \textbf{pussy} & peckpalitchoke & standwithhongkong & bigolive \\
        & & \textbf{afd} & \textbf{blackpink} & \textbf{sex} & vixx & hongkong & 10tv \\
        & & \textbf{cdnpoli} & \textbf{pcas} & \textbf{horny} & wemadeit & bigil & mixch.tv(ja) \\
        & & \textbf{elxn43} & \textbf{exo} & \textbf{porn} & mayward & lebanon & twitch \\
        \bottomrule
        \toprule
        \multirow{11}{*}{\rotatebox[origin=c]{90}{sample set}} & & \textbf{SC1} & \textbf{SC2} & \textbf{SC3} & SC4 & SC5 & SC6 \\
        & size & \textbf{1,880,247} & \textbf{823,232} & \textbf{551,219} & 822,436 & 549,589 & 805,852 \\
        & \#users & \textbf{1,600,579} & \textbf{767,183} & \textbf{446,303} & 686,609 & 465,339 & 688,922 \\
        & \#hashtags & \textbf{279,668} & \textbf{56,049} & \textbf{104,916} & 135,827 & 84,250 & 116,930 \\
        & avg. degree & \textbf{5.58} & \textbf{5.75} & \textbf{14.98} & 3.06 & 3.51 & 3.28 \\
        & category & \textbf{politics} & \textbf{Korean pop} & \textbf{cyberbullying} & mixed & mixed & mixed \\
        \cmidrule{2-8}
        & \multirow{5}{*}{hashtags} & \textbf{ps4live} & \textbf{bts} & \textbf{gay} & mixch.tv(ja) & bigolive & Idolish7(ja) \\
        & & \textbf{10tv} & \textbf{mamavote} & \textbf{pussy} & bigil & kamleshtiwari & reunion \\
        & & \textbf{brexit} & \textbf{blackpink} & \textbf{sex} & peckpalitchoke(th) & bb13 & Idolish7(ja) \\
        & & \textbf{afd} & \textbf{pcas} & \textbf{horny} & reality\_about\_islam(hi) & biggboss13 & vixx  \\
        & & \textbf{demdebate} & \textbf{bts(ko)} & \textbf{porn} & doki.live(ja) & execution\_rajeh\_mahmoud(ar) & vixx(ko) \\
        \bottomrule
    \end{tabular}
    }
    \caption{Statistics and the most used 5 hashtags in the 6 clusters of the user-hashtag bipartite graph.
    Three complete clusters maintain their structure in the sample set (\textbf{boldfaced}).
    The language code within brackets is the original language for the hashtag. ja: Japanese; ko: Korean; th: Thai; hi: Hinda; ar: Arabic.}
    \label{table:clustering}
\end{table*}

Clustering techniques are often used to detect communities in such bipartite graphs.
We apply spectral clustering~\cite{stella2003multiclass} on the user-hashtag bipartite graph, with the number of clusters set at 6.
The resulted clusters are summarized in \Cref{table:clustering}, together with the most used 5 hashtags and a manually-assigned category.
Apart from the cyberbullying keywords, there are significant amount of hashtags related to politics, live streaming, and Korean pop culture, which are considered as some of the most discussed topics on Twitter.
We further quantify how the clusters traverse from the complete set to the sample set in \Cref{fig:bipartite-cluster}.
Three of the complete clusters (CC1, CC2, and CC3) are maintained in the sample set (mapping to SC1, SC2, and SC3 respectively), since more than half of the entities preserve.
The remaining three complete clusters disperse.
Investigating the statistics for the complete clusters, the preserved ones have a larger average weighted degree, meaning more tweets between the users and hashtags in these clusters.
Another notable observation is that albeit the entities traverse to the sample clusters differently, all complete clusters have similar missing rates (28\% to 34\%).
It suggests that Twitter data sampling impacts the community structure.
Denser structures are more resilient to sampling.


\subsection{User-user retweet network}
\label{ssec:retweet}


\begin{figure}[tbp]
    \centering
    \includegraphics[width=0.8\columnwidth]{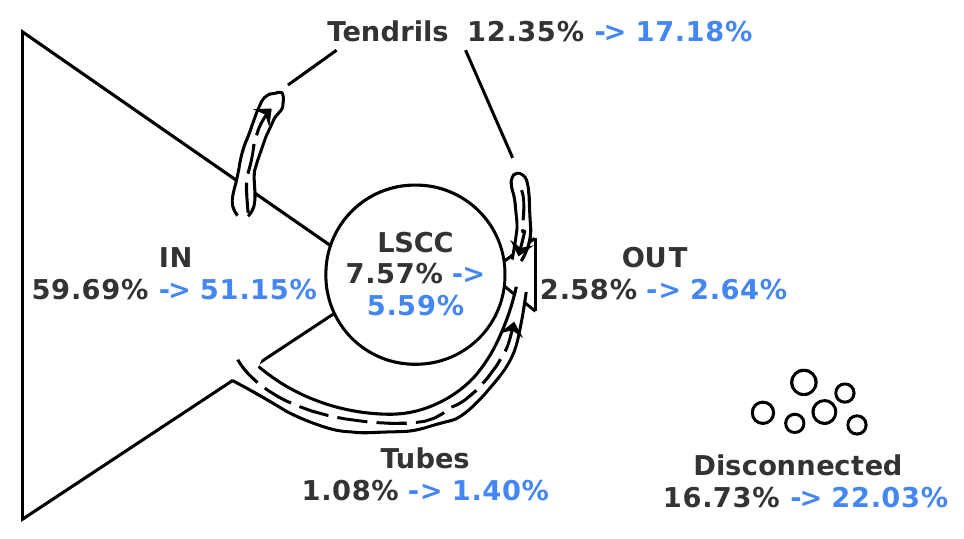} 
	\caption{Visualization of bow-tie structure in complete set. The black number indicates the relative size of component in the complete set, \textcolor{blue}{blue} number indicates the relative size in the sample set.
    }
	\label{fig:measure-bowtie}
\end{figure}

Retweet network describes the information sharing between users.
We build a user-user retweet network by following the ``@RT'' relation..
Each node is a user, and each edge is a directed link weighted by the number of retweets between two users.
The user-user retweet network has been extensively investigated in literature~\cite{sadikov2011correcting,morstatter2013sample,gonzalez2014assessing}.

We choose to characterize the retweet network using the bow-tie structure.
Initially proposed to measure the World Wide Web~\cite{broder2000graph}, the bow-tie structure was also used to measure the QA community~\cite{zhang2007expertise} or YouTube video networks~\cite{wu2019estimating}.
The bow-tie structure characterizes a network into 6 components:
(a) the largest strongly connected component (LSCC) as the central part;
(b) the IN component contains nodes pointing to LSCC but not reachable from LSCC;
(c) the OUT component contains nodes that can be reached by LSCC but not pointing back to LSCC;
(d) the Tubes component connects the IN and OUT components;
(e) the Tendrils component contains nodes pointing from In component or pointing to OUT component;
(f) the Disconnected component includes nodes not in the above 5 components.
\Cref{fig:measure-bowtie} visualizes the bow-tie structure of the user-user retweet network, alongside with the relative size for each component in the complete and sample sets.
The LSCC and IN components, which make up the majority part of the bow-tie, reduce the most in both absolute size and relative ratio due to sampling.
OUT and Tubes are relatively small in both complete and sample sets.
Tendrils and disconnected components enlarge 39\% and 32\% after sampling.

\begin{figure}[tbp]
    \centering
    \includegraphics[width=.99\columnwidth]{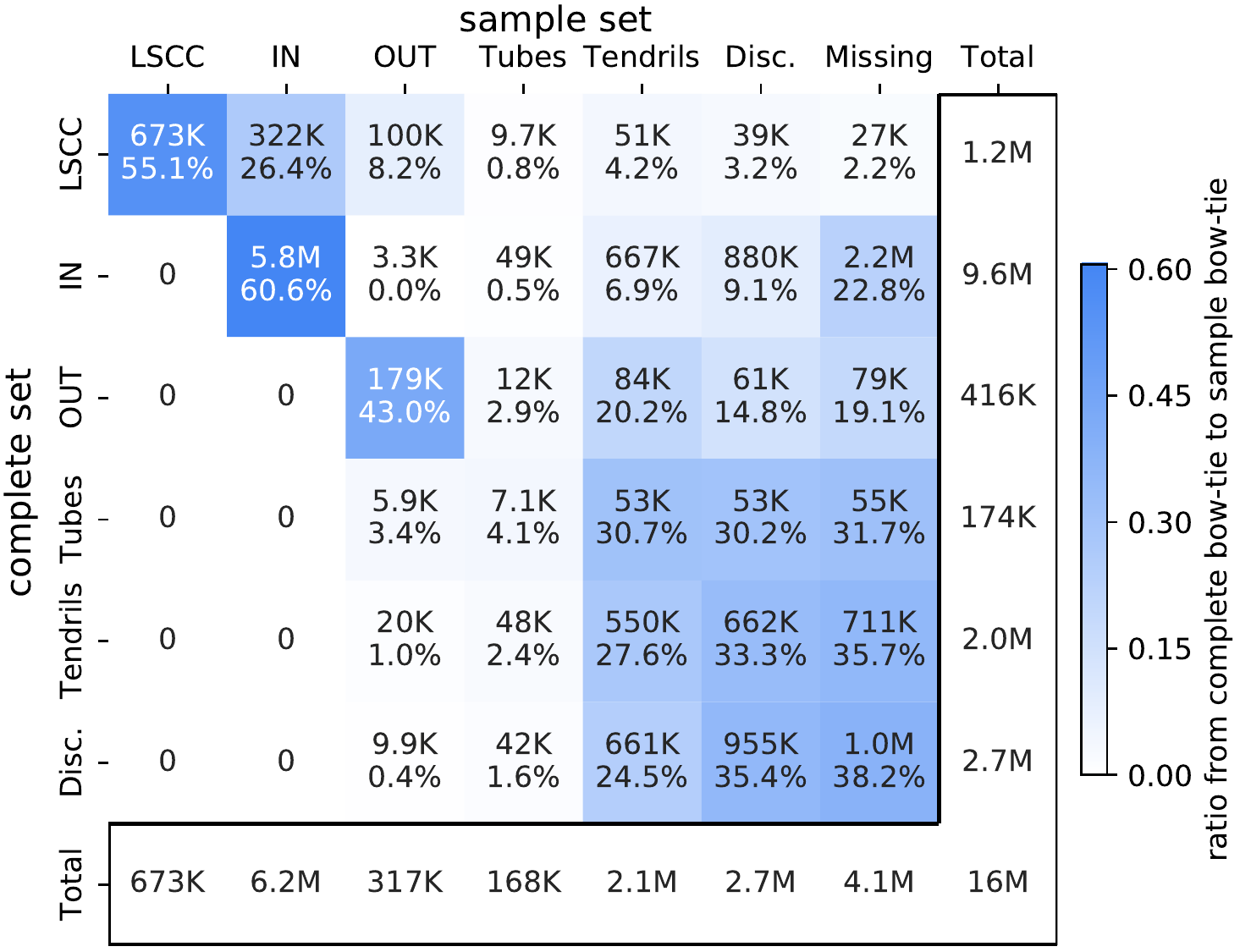}
	\caption{The change of bow-tie components from complete set to sample set.
	Each cell denotes the volume (top) and the ratio (bottom) of users that traverse from a component in complete set to a component in sample set.}
	\label{fig:measure-bowtie-stats}
\end{figure}

\Cref{fig:measure-bowtie-stats} shows the node flow of each components from the complete set to the sample set.
About a quarter of LSCC component shift to the IN component.
For the OUT, Tubes, Tendrils, and Disconnected components, 20\% to 31\% nodes move into the Tendrils component, resulting in a slight increase of absolute size for Tendrils.
Most notably, nodes in the LSCC has a much smaller chance of missing (2.2\%, other components are with 19\% to 38\% missing rates).



\section{Impacts on retweet cascades}
\label{sec:cascade}

Information diffusion is perhaps the most studied social phenomenon on Twitter.
A retweet cascade consists of two parts: a root tweet and its subsequent retweets.
A number of models have been proposed for modeling and predicting retweet cascades~\cite{zhao2015seismic,mishra2016feature,martin2016exploring}.
However, these usually make the assumption of observing all the retweets in cascades.
In this section, we analyze the impacts of Twitter sampling on retweet cascades and identify risks for existing models.
We first construct cascades without missing tweets from the complete set.
Next, we measure the sampling effects for some commonly used features in modeling retweet cascades, e.g., inter-arrival time and potential reach.

\header{Constructing complete cascades.}
When using the filtered streaming API, if a root tweet is observed, the API should return all its retweets.
This is because the API also tracks the keywords in the \texttt{retweeted\_status} field of a tweet (i.e., the root tweet), which allows us to construct a set of complete cascades from the complete set.
In the sample set, both the root tweet and any of its retweets could be missing.
If the root tweet is missing, we miss the entire cascade.
If some retweets are missing, we observe a partial cascade.
\Cref{table:cascade} lists the obtained cascades in the complete and the sample sets.
Notably, there are 3M cascades in the complete set, but only 1.17M in the sample set (38.85\%), out of which only 508k (16.88\%) cascades are complete and their sizes are relatively small (i.e., they don't miss any retweet, max cascade size: 23, mean size: 1.37).
Prior literature~\cite{zhao2015seismic} often concentrates on retweet cascades with more than 50 retweets.
There are 99,952 such cascades in the complete set, but only 29,577 in the sample set, out of which none is complete.

\begin{figure}[tbp]
    \centering
    \includegraphics[width=.99\columnwidth]{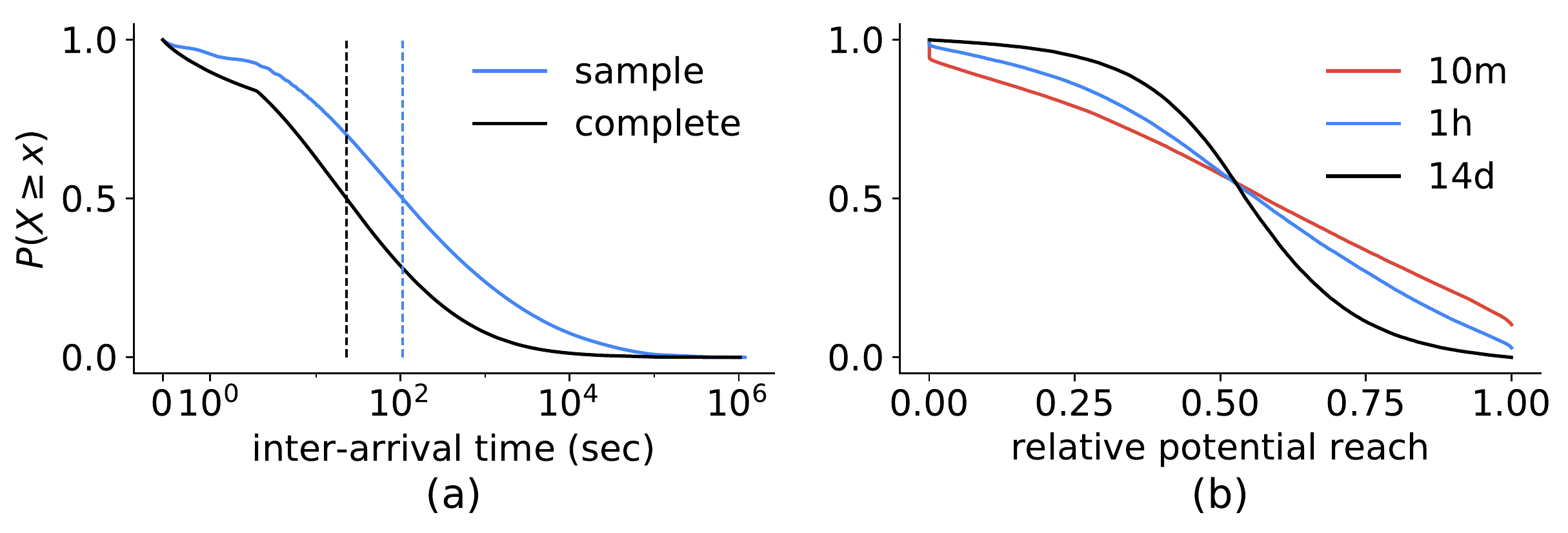}
	\caption{CCDFs of \textbf{(a)} inter-arrival time and \textbf{(b)} relative potential reach.}
	\label{fig:measure-cascade}
\end{figure}

\header{Inter-arrival time.} One line of work models the information diffusion as point processes~\cite{zhao2015seismic,mishra2016feature}.
These models use a memory kernel as a function of the time gap $\Delta t$ between two consecutive events, which is also known as inter-arrival time.
\Cref{fig:measure-cascade}(a) plots the CCDFs of inter-arrival times in the complete and the sample sets.
The distribution shifts right, towards larger values.
This is expected as the missing tweets increase the time gap between two observed tweets.
The median inter-arrival time is 22.9 seconds in the complete set (black dashed line), meaning 50\% retweets happen within 23 seconds from last retweet.
After sampling, the median increases almost 5-fold to 105.7 seconds (\textcolor{blue}{blue} dashed line).
For research that uses tweet inter-arrival time, this presents the risk of miss-calibrating models and of underestimating the virality of the cascades.


\begin{table}
    \centering
    \resizebox{.99\columnwidth}{!}{
    \begin{tabular}{rrrr}
        \toprule
         & complete & sample & ratio \\
        \midrule
        \#cascades & 3,008,572 & 1,168,896 & 38.9\% \\
        \#cascades (${\geq}$50 retweets) & 99,952 & 29,577 & 29.6\% \\
        avg. retweets per cascade & 15.6 & 11.0 & 70.2\% \\
        med. inter-arrival time (s) & 22.9 & 105.7 & 461.6\% \\
        \bottomrule
    \end{tabular}
    }
    \caption{Statistics of cascades in \cb dataset.}
    \label{table:cascade}
\end{table}


\header{Potential reach.}
Online influence is another well-studied phenomenon on Twitter, and one of its proxies is the number of followers of a user.
We define potential reach as the total number of all observed retweeters' followers.
This approximates the size of the potential audience for the root tweet.
We compute the relative potential reach as the ratio of potential reach in the sample cascade against that in the complete cascade, and we plot the CCDFs in \Cref{fig:measure-cascade}(b).
When observing cascades for as much as 14 days, 50\% of the cascades have the relative potential reach below 0.544.
This indicates that when using the sampled Twitter data, researchers can severely underestimate the size of the potential audience.
Another common setting is to perform early prediction, i.e., after observing 10 minutes or 1 hour of each retweet cascade.
\Cref{fig:measure-cascade}(b) shows that the relative potential reach is more evenly distribution for shorter time windows -- 21.0\% cascades have relative potential reach below 0.25 and 33.7\% cascades above 0.75 within 10 minutes span -- comparing to the observation over 14 days (5.1\% and 11.3\%, respectively).





\section{Conclusion}
\label{sec:conclusion}

This work presents a set of in-depth measurements on the effects of Twitter data sampling.
We validate that Twitter rate limit messages closely approximate the volume of missing tweets.
Across different timescales (hour, minute, second, millisecond), we find that the sampling rates have distinct temporal variations at each scale.
We show the effects of sampling across different subjects (entities, networks, cascades), which may in turn distort the results and interpretations of measurement and modeling studies.
For counting statistics such as number of tweets per user and per hashtag, we find that the Bernoulli process with a uniform rate is a reasonable approximation for Twitter data sampling.
We also show how to estimate ground-truth statistics in the complete data by using only the sample data.

\header{Limitations.}
These observations in this paper apply to current Twitter APIs (as of 2020-03) and are subject to the changes of Twitter's proprietary sampling mechanisms.
We are aware of that Twitter plans to release a new set of APIs in near future.
Consistent with the current streaming APIs, the rate limit threshold for the new APIs is also set to 50 tweets per second~\cite{twitterrate2}.
Therefore, we believe the observations of this paper will hold.

\header{Practical implications and future work.}
This work calls attention to the hidden biases in social media data.
We have shown effective methods for estimating ground-truth statistics, which allows researchers to mitigate the risks in their datasets without collecting the complete data.
Our research also provides methods and toolkits for collecting sampled and complete data streams on Twitter.
Our findings provide foundations to many other research topics using sampled data, such as community detection and information diffusion algorithms that are robust to data subsampling.
Future works include measuring a larger set of activity and network measurements under data sampling, generalizing the results of this work to other social media platforms and data formats, and quantifying the robustness of existing network and diffusion models against data sampling. 


\section*{Acknowledgments}
This work is supported in part by AOARD Grant 19IOA078, Australian Research Council Project DP180101985, and the National eResearch Collaboration Tools and Resources (Nectar).
Marian-Andrei Rizoiu's time is partially supported by a Facebook content policy award.





\bibliography{twitter-sampling-ref}
\bibliographystyle{aaai}


\clearpage
\appendix
\onecolumn
\normalsize

Accompanying the submission \textit{Variation across Scales: Measurement Fidelity under Twitter Data Sampling}.

\section{Twitter data in ICWSM papers (2015-2019)}
\label{sec:si_icwsm}

82 (31\%) out of 265 ICWSM full papers used Twitter data from 2015 to 2019.
Twitter search API has been used 25 times, sampled stream 12 times, filtered stream 18 times, firehose 8 times.
12 papers used multiple Twitter APIs for data collection.
7 papers did not clearly specify their Twitter API choices.

\begin{table*}[!htbp]
    \centering
    \caption{Year 2015. 20 out of 64 papers used Twitter data. search API: 4; sampled stream: 5; filtered stream: 3; firehose: 1; unspecified: 4; multiple APIs: 3.}
    \label{table:2015}
    \footnotesize
    \begin{tabular}{>{\centering\arraybackslash}p{0.2cm}%
    >{\raggedright\arraybackslash}p{7cm}%
    >{\raggedright\arraybackslash}p{2.3cm}%
    >{\raggedright\arraybackslash}p{6.4cm}%
    }
        \toprule
        Id & Paper & APIs & Notes\\
        \midrule
        1 & Audience analysis for competing memes in social media & search & searched keywords ``Russia'', ``meteor'', ``Fox'', and ``Obama'' \\
        2 & Making use of derived personality: The case of social media ad targeting & filtered & mention at least one term related to NYC and one term related to traveling \\
        3 & The many shades of anonymity: Characterizing anonymous social media content & unspecified, possibly sampled & 500 random publicly available tweets \\
        4 & On analyzing hashtags in Twitter & sampled; search & 10M messages crawled in December 2013; 200 tweets for each hashtag in our original dataset \\
        5 & WhichStreams: A dynamic approach for focused data capture from large social media & sampled; filtered & 5000 users first to use one of the keywords ``Obama'', ``Romney'' or ``\#USElections'' \\
        6 & Characterizing silent users in social media communities & filtered & all tweets of 140,851 Singapore-based users and 126,047 Indonesia-based users \\
        7 & Predicting user engagement on Twitter with real-world events & firehose & nearly 2.7 billion English tweets during August of 2014 \\
        8 & Geolocation prediction in Twitter using social networks: A critical analysis and review of current practice & sampled & 10\% sampled stream \\
        9 & Characterizing information diets of social media users & sampled & 500 randomly selected tweets from Twitter's 1\% random sample \\
        10 & Degeneracy-based real-time sub-event detection in Twitter stream & unspecified, possibly filtered & several football matches that took place during the 2014 FIFA World Cup in Brazil, between the 12th of June and the 13th of July 2014 \\
        11 & CREDBANK: A large-scale social media corpus with associated credibility annotations & sampled & 1\% random sample \\
        12 & Understanding musical diversity via online social media & search & collected U.S. Twitter users who share their Last.fm accounts, then we collected all publicly available tweets \\
        13 & Smelly maps: The digital life of urban smellscapes & unspecified & collected 5.3M tweets during year 2010 and from October 2013 to February 2014 \\
        14 & Project recommendation using heterogeneous traits in crowdfunding & search & retrieving tweets containing URLs that begin with \url{http://kck.st} \\
        15 & Don't let me be \#misunderstood: Linguistically motivated algorithm for predicting the popularity of textual memes & sampled & approximately 15\% of the Twitter stream in six month period \\
        16 & SEEFT: Planned social event discovery and attribute extraction by fusing Twitter and web content & unspecified, possibly search & querying Twitter API with 3 event types, namely concerts, conferences, and festivals \\
        17 & A bayesian graphical model to discover latent events from Twitter & sampled & 1\% sampled stream and 10\% sampled stream \\
        18 & Patterns in interactive tagging networks & sample; search & randomly sampled 1 million seed users from sample streams on Dec 2014; following network starting from the same 1 million seed users \\
        19 & Hierarchical estimation framework of multi-label classifying: A case of tweets classifying into real life aspects & search & collected 2,390,553 tweets posted from April 15, 2012 to August 14, 2012, each of which has ``Kyoto'' as the Japanese location information \\
        20 & The lifecyle of a Youtube video: Phases, content and popularity & filtered & tweets containing keyword ``youtube'' OR (``youtu'' AND ``be'') \\
        \bottomrule
    \end{tabular}
\end{table*}

\clearpage

\begin{table*}[!htbp]
    \centering
    \caption{Year 2016. 23 out of 52 papers used Twitter data. search API: 10; sampled stream: 2; filtered stream: 5; firehose: 3; unspecified: 1; multiple APIs: 2.}
    \label{table:2016}
    \footnotesize
    \begin{tabular}{>{\centering\arraybackslash}p{0.2cm}%
    >{\raggedright\arraybackslash}p{7cm}%
    >{\raggedright\arraybackslash}p{2.3cm}%
    >{\raggedright\arraybackslash}p{6.4cm}%
    }
        \toprule
        Id & Paper & APIs & Notes\\
        \midrule
        1 & Are you charlie or ahmed? Cultural pluralism in charlie hebdo response on Twitter & search & \#JeSuisCharlie, \#JeSuisAhmed, and \#CharlieHebdo -- from 2015-01-07 to 2015-01-28 \\
        2 & When a movement becomes a party: Computational assessment of new forms of political organization in social media & filtered & extracted 373,818 retweets of tweets that (1) were created by, (2) were retweeted by, or (3) mentioned a user from the list \\
        3 & Journalists and Twitter: A multidimensional quantitative description of usage patterns & search & contained 5,358 accounts of journalists and news organizations, crawled all their 13,140,449 public tweets \\
        4 & Social media participation in an activist movement for racial equality & filtered & \#ferguson, \#BlueLivesMatter, \#BlackLivesMatter, \#AllLivesMatter, \#Baltimore, \#BaltimoreRiots, \#BaltimoreUprising, and \#FreddieGray \\
        5 & Understanding communities via hashtag engagement: A clustering based approach & firehose & tweets from all English language Twitter users in U.S. that used a hashtag at least once during the 30 day study period starting Jan 15, 2015 \\
        6 & Investigating the observability of complex contagion in empirical social networks & filtered; search & 45 Nigerian cities with populations of 100K or more using a radius from 25 to 40 miles; collected tweets from the timelines of selected users\\
        7 & Dynamic data capture from social media streams: A contextual bandit approach & sampled; filtered & leverage sampled stream to discover unknown users; filtered stream for realtime data of the subset users\\
        8 & On unravelling opinions of issue specific-silent users in social media & search & asked the Twitter users to provide their Twitter screen names so as to crawl their Twitter data  \\
        9 & Distinguishing between topical and non-topical information diffusion mechanisms in social media & search & a dataset that is nearly complete and contains all public tweets produced by users until Sep 2009 and a snapshot of the social graph in Sep 2009 \\
        10 & TweetGrep: Weakly supervised joint retrieval and sentiment analysis of topical tweets & search & the queries are issued to the Twitter Search Web Interface via a proxy that we developed  \\
        11 & What the language you tweet says about your occupation & search & we download these users' 3,000 most recent tweets \\
        12 & TiDeH: Time-dependent hawkes process for predicting retweet dynamics & firehose & SEISMIC dataset by Zhao et al. 2015 \\
        13 & Emotions, demographics and sociability in Twitter interactions & search & collect tweets from an area that included Los Angeles, then collect all (timeline) tweets from subset users \\
        14 & Analyzing personality through social media profile picture choice & search & we have collected up to 3,200 most recent tweets for each user \\
        15 & Cross social media recommendation & unspecified, possibly sampled & corpora were sampled between 2012/09/17 and 2012/09/23\\
        16 & Understanding anti-vaccination attitudes in social media & firehose & a manual examination of 1000 Twitter posts, then snowball sampling from a sample of firehose \\
        17 & Twitter's glass ceiling: The effect of perceived gender on online visibility & sampled & 10\% sampled stream \\
        18 & Mining pro-ISIS radicalisation signals from social media users & search & Twitter user timeline of 154K users \\
        19 & Predictability of popularity: Gaps between prediction and understanding & sampled &  URLs tweeted by 737k users for three weeks of 2010 \\
        20 & Theme-relevant truth discovery on Twitter: An estimation theoretic approach & search & collected through Twitter search API using query terms and specified geographic regions related to the events\\
        21 & \#PrayForDad: Learning the semantics behind why social media users disclose health information & filtered & collect tweets in English and published in the contiguous United States during a four-month window in 2014 \\
        22 & Your age is no secret: Inferring microbloggers' ages via content and interaction analysis & filtered &  record all the tweets which contain one of the keywords ``happy \textit{y}th birthday'' with y ranging from 14 to 70 \\
        23 & EigenTransitions with hypothesis testing: The anatomy of urban mobility & filtered & collected geo-tagged Tweets generated within the area covering New York City and Pittsburgh from Jul 15, 2013 to Nov 09, 2014. \\
        \bottomrule
    \end{tabular}
\end{table*}

\clearpage

\begin{table*}[!htbp]
    \centering
    \caption{Year 2017. 9 out of 50 papers used Twitter data. search API: 3; filtered stream: 3; multiple APIs: 3.}
    \label{table:2017}
    \footnotesize
    \begin{tabular}{>{\centering\arraybackslash}p{0.2cm}%
    >{\raggedright\arraybackslash}p{7cm}%
    >{\raggedright\arraybackslash}p{2.3cm}%
    >{\raggedright\arraybackslash}p{6.4cm}%
    }
        \toprule
        Id & Paper & APIs & Notes\\
        \midrule
        1 & Who makes trends? Understanding demographic biases in crowdsourced recommendations & sampled; search & 1\% random sample; queried search API every 5 minutes and collected all topics which became trending in US \\
        2 & \#NotOkay: Understanding gender-based violence in social media & sampled; filtered & 1\% random sample;  collect tweets containing indicated hashtags from Oct 26 to Nov 26, 2016 \\
        3 & Online popularity under promotion: Viral potential, forecasting, and the economics of time & filtered & tweets containing keyword ``youtube'' OR (``youtu'' AND ``be'')  \\
        4 & Examining the alternative media ecosystem through the production of alternative narratives of mass shooting events on Twitter & filtered & tracked ``shooter, shooting, gunman, gunmen, gunshot, gunshots, shooters, gun shot, gun shots, shootings'' between Jan 1 and Oct 5, 2016 \\
        5 & State of the geotags: Motivations and recent changes & filtered & selected all coordinate-geotagged tweets within 0.2 degrees latitude and longitude from Pittsburgh \\
        6 & Online human-bot interactions: Detection, estimation, and characterization & search & collected the most recent tweets produced by those accounts \\
        7 & Identifying effective signals to predict deleted and suspended accounts on Twitter across languages & sampled; search & 1\% random sample; batches of 100 unique users were queried against the public Twitter API \\
        8 & Adaptive spammer detection with sparse group modeling & search & crawled a Twitter dataset from July 2012 to September 2012 via the Twitter Search API \\
        9 & Wearing many (social) hats: How different are your different social network personae? & search & 76\% of About.me users in our dataset have linked their profiles to their alternate account in Twitter \\
        \bottomrule
    \end{tabular}
\end{table*}

\begin{table*}[!htbp]
    \centering
    \caption{Year 2018. 13 out of 48 papers used Twitter data. search API: 2; sampled stream: 3; filtered stream: 4; firehose: 2; multiple APIs: 2.}
    \label{table:2018}
    \footnotesize
    \begin{tabular}{>{\centering\arraybackslash}p{0.2cm}%
    >{\raggedright\arraybackslash}p{7cm}%
    >{\raggedright\arraybackslash}p{2.3cm}%
    >{\raggedright\arraybackslash}p{6.4cm}%
    }
        \toprule
        Id & Paper & APIs & Notes\\
        \midrule
        1 & Peer to peer hate: Hate speech instigators and their targets & sampled; search & 1\% random sample;  we use search API to fetch tweet traces of users \\
        2 & Characterizing audience engagement and assessing its impact on social media disclosures of mental illnesses & search& obtain the user list of who has retweeted the disclosers during this period of analysis \\
        3 & Facebook versus Twitter: Cross-platform differences in self-disclosure and trait prediction & search & we collected participants' social media posts \\
        4 & Can you verifi this? Studying uncertainty and decision-making about misinformation using visual analytics & filtered & collected 103,248 tweets posted by these 178 accounts along with account metadata from May 23, 2017 to June 6, 2017 \\
        5 & Using longitudinal social media analysis to understand the effects of early college alcohol use & firehose & extract 639k tweets that match these keywords in August-December 2010 in our organization's archive of the Twitter firehose \\
        6 & Modeling popularity in asynchronous social media streams with recurrent neural networks & filtered; firehose & tweets containing keyword ``youtube'' OR (``youtu'' AND ``be''); SEISMIC dataset by Zhao et al. 2015 \\
        7 & The effect of extremist violence on hateful speech online & sampled & 10\% random sample\\
        8 & You are your metadata: Identification and obfuscation of social media users using metadata information & sampled & random sample of the tweets posted between October 2015 and January 2016 \\
        9 & \#DebateNight: The role and influence of socialbots on Twitter during the first 2016 U.S. presidential debate & firehose & Twitter discussions that occurred during the 1st 2016 U.S presidential debate between Hillary Clinton and Donald Trump \\
        10 & Ecosystem or echo-system? Exploring content sharing across alternative media domains & filtered & tracked various keyword terms related to the Syrian conflict including geographic terms of affected areas \\
        11 & COUPLENET: Paying attention to couples with coupled attention for relationship recommendation & filtered & collected tweets with emojis contains the keyword ``heart'' in its description \\
        12 & Beyond views: Measuring and predicting engagement in online videos & filtered & tweets containing keyword ``youtube'' OR (``youtu'' AND ``be'') \\
        13 & Understanding web archiving services and their (mis)use on social media & sampled & 1\% random sample \\
        \bottomrule
    \end{tabular}
\end{table*}

\clearpage

\begin{table*}[!htbp]
    \centering
    \caption{Year 2019. 17 out of 51 papers used Twitter data. search API: 6; sampled stream: 2; filtered stream: 3; firehose: 2; unspecified: 2; multiple APIs: 2.}
    \label{table:2019}
    \footnotesize
    \begin{tabular}{>{\centering\arraybackslash}p{0.2cm}%
    >{\raggedright\arraybackslash}p{7cm}%
    >{\raggedright\arraybackslash}p{2.3cm}%
    >{\raggedright\arraybackslash}p{6.4cm}%
    }
        \toprule
        Id & Paper & APIs & Notes\\
        \midrule
        1 & Linguistic cues to deception: Identifying political trolls on social media & firehose & a list of 2,752 Russian troll accounts, then collected all of the trolls' discussions \\
        2 & Tweeting MPs: Digital engagement between citizens and members of parliament in the UK & search & we fetched all the users (?4.28 Million) who follow MPs and also the users that MPs followed (869K) \\
        3 & View, like, comment, post: Analyzing user engagement by topic at 4 levels across 5 social media platforms for 53 news organizations & filtered & collecting all posts from a news organization \\
        4 & A large-scale study of ISIS social media strategy: Community size, collective influence, and behavioral impact & firehose &  a large dataset of 9.3 billion tweets representing all tweets generated in the Arabic language in 2015 through full private access to the Twitter firehose \\
        5 & Who should be the captain this week? Leveraging inferred diversity-enhanced crowd wisdom for a fantasy premier league captain prediction & unspecified & collected their soccer related tweets by scraping Twitter user timelines (for a total 4,299,738 tweets) \\
        6 & Multimodal social media analysis for gang violence prevention & search &  we scraped all obtainable tweets from this list of 200 users in February 2017 \\
        7 & Hot streaks on social media & search & we obtained all tweets, followers, and retweeters of all tweets using the Twitter REST API \\
        8 & Understanding and measuring psychological stress using social media & search & 601 active users who completed the survey \\
        9 & Studying cultural differences in emoji usage across the east and the west & sampled & 10\% random sample \\
        10 & What Twitter profile and postedImages reveal about depression and anxiety & search & downloaded the 3200 most recent user tweets for each user, leading to a data set of 5,547,510 tweets, out of which 700,630 posts contained images and 1 profile image each across 3498 users \\
        11 & Polarized, together: Comparing partisan support for Trump's tweets using survey and platform-based measures & sampled; search & collecting a large sample of Twitter users (approximately 406M) who sent one or more tweets that appeared in the Twitter Decahose from January 2014 to August 2016; select from this set the  approximately 322M accounts that were still active in March 2017 \\
        12 & Race, ethnicity and national origin-based discrimination in social media and hate crimes across 100 U.S. cities & sampled & 1\% sample of Twitter's public stream from January 1st, 2011 to December 31st,  2016 \\
        13 & A social media study on the effects of psychiatric medication use & sampled; filtered & public English posts mentioning these drugs between January 01, 2015 and December 31, 2016 \\
        14 & SENPAI: Supporting exploratory text analysis through semantic\&syntactic pattern inspection& filtered & gathered a dataset of Twitter messages from 103 professional journalists and bloggers who work in the field of American Politics \\
        15 & Empirical analysis of the relation between community structure and cascading retweet diffusion & search & we used the Search API and collected Japanese tweets using the query q=RT, lang=ja \\
        16 & Measuring the importance of user-generated content to search engines & unspecified & a row of three cards with one tweet each. Google obtains the tweets either from Twitter's search (a SearchTweetCarousel) or a single user (a UserTweetCarousel) \\
        17 & Detecting journalism in the age of social media:Three experiments in classifying journalists on Twitter & filtered & tracking a set of event-related keywords and hashtags \\
        \bottomrule
    \end{tabular}
\end{table*}

\clearpage

\section{Constructing the complete data streams}
\label{sec:si_crawler}

To collect the complete streams, we split the same set of keywords into multiple streaming clients.
The \textsc{Cyberbullying} and \textsc{YouTube} datasets are respectively crawled by 8 and 12 clients based on different combinations of keywords and languages.
We carefully choose the split criteria to ensure that each client would not be rate limited by much ($\bar{\rho}{>}99\%$ for all subcrawlers).
Table \ref{table:conf_cb} and Table \ref{table:conf_yt} list the statistics for all streaming clients of the two datasets.

\begin{itemize}[leftmargin=*]
  \item 25 keywords for the \textsc{Cyberbullying} dataset: \texttt{nerd}, \texttt{gay}, \texttt{loser}, \texttt{freak}, \texttt{emo}, \texttt{whale}, \texttt{pig}, \texttt{fat}, \texttt{wannabe}, \texttt{poser}, \texttt{whore}, \texttt{should}, \texttt{die}, \texttt{slept}, \texttt{caught}, \texttt{suck}, \texttt{slut}, \texttt{live}, \texttt{afraid}, \texttt{fight}, \texttt{pussy}, \texttt{cunt}, \texttt{kill}, \texttt{dick}, \texttt{bitch}
  
  \item rules for the \textsc{YouTube} dataset: ``\texttt{youtube}'' \texttt{OR} (``\texttt{youtu}'' \texttt{AND} ``\texttt{be}'')
  
  \item 66 language codes on Twitter: en, es, ja, ko, und, ar, pt, de, tl, fr, cs, it, vi, in, tr, pl, ru, sr, th, el, nl, hi, zh, da, ro, is, no, hu, fi, lv, et, bg, ht, uk, lt, cy, ka, ur, sv, ta, sl, iw, ne, fa, am, te, km, ckb, hy, eu, bn, si, my, pa, ml, gu, kn, ps, mr, sd, lo, or, bo, ug, dv, ca
\end{itemize}

\begin{table*}[!htbp]
    \centering
    \caption{Subcrawler configurations for \textsc{Cyberbullying} dataset. all$\setminus$en is all languages excluding ``en''.}
    \label{table:conf_cb}
    \footnotesize
    \begin{tabular}{r>{\centering\arraybackslash}p{3cm}rrrrr}
        \toprule
        Id & Keywords & Languages & \#collected tweets & \#rate limit & \#est. missing tweets & sampling rate\\
        \midrule
        1 & \texttt{should} & en & 29,647,814 & 1,357 & 7,324 & 99.98\% \\
        2 & \texttt{should} & all$\setminus$en & 801,904 & 0 & 0 & 100.00\% \\
        3 & \texttt{live} & en & 16,526,226 & 1,273 & 25,976 & 99.84\% \\
        4 & \texttt{live} & all$\setminus$en & 7,926,325 & 233 & 7,306 & 99.91\% \\
        5 & \texttt{kill}, \texttt{fight}, \texttt{poser}, \texttt{nerd}, \texttt{freak}, \texttt{pig} & all & 15,449,973 & 16 & 108 & 100.00\% \\
        6 & \texttt{dick}, \texttt{suck}, \texttt{gay}, \texttt{loser}, \texttt{whore}, \texttt{cunt} & all & 13,164,053 & 15 & 125 & 100.00\% \\
        7 & \texttt{pussy}, \texttt{fat}, \texttt{die}, \texttt{afraid}, \texttt{emo}, \texttt{slut} & all & 21,333,866 & 89 & 1,118 & 99.99\% \\
        8 & \texttt{bitch}, \texttt{wannabe}, \texttt{whale}, \texttt{slept}, \texttt{caught} & all & 14,178,366 & 64 & 666 & 100.00\% \\
        complete & subcrawlers 1-8 & all & 114,488,537 & 3,047 & 42,623 & 99.96\% \\
        sample & all 25 keywords & all & 60,400,257 & 1,201,315 & 54,175,503 & 52.72\% \\
        \bottomrule
    \end{tabular}
\end{table*}

\begin{table*}[!htbp]
    \centering
    \caption{Subcrawler configurations for \textsc{YouTube} dataset.}
    \label{table:conf_yt}
    \footnotesize
    \begin{tabular}{r>{\centering\arraybackslash}p{3cm}rrrrr}
        \toprule
        Id & Keywords & Languages & \#collected tweets & \#rate limit & \#est. missing tweets & sampling rate \\
        \midrule
        1 & ``\texttt{youtube}'' & en & 10,312,498 & 323 & 3,582 & 99.97\% \\
        2 & ``\texttt{youtube}'' & ja & 6,620,927 & 118 & 3,211 & 99.95\% \\
        3 & ``\texttt{youtube}'' & ko & 714,992 & 36 & 1,339 & 99.81\% \\
        4 & ``\texttt{youtube}'' & es & 2,106,474 & 0 & 0 & 100.00\% \\
        5 & ``\texttt{youtube}'' & und & 1,418,710 & 0 & 0 & 100.00\% \\
        6 & ``\texttt{youtube}'' & all$\setminus$\{en,ja,ko,es,und\} & 5,264,150 & 20 & 169 & 100.00\% \\
        7 & ``\texttt{youtu}'' \texttt{AND} ``\texttt{be}'' & en & 11,188,872 & 530 & 10,328 & 99.91\% \\
        8 & ``\texttt{youtu}'' \texttt{AND} ``\texttt{be}'' & ja & 8,389,060 & 619 & 9,657 & 99.89\% \\
        9 & ``\texttt{youtu}'' \texttt{AND} ``\texttt{be}'' & ko & 4,560,793 & 1,193 & 43,584 & 99.05\% \\
        10 & ``\texttt{youtu}'' \texttt{AND} ``\texttt{be}'' & es & 2,271,712 & 27 & 829 & 99.96\% \\
        11 & ``\texttt{youtu}'' \texttt{AND} ``\texttt{be}'' & und & 2,856,415 & 37 & 1,556 & 99.95\% \\
        12 & ``\texttt{youtu}'' \texttt{AND} ``\texttt{be}'' & all$\setminus$\{en,ja,ko,es,und\} & 7,351,671 & 158 & 2,800 & 99.96\% \\
        complete & subcrawlers 1-12 & all & 53,557,950 & 3,061 & 77,055 & 99.86\% \\
        sample & ``\texttt{youtube}'' \texttt{OR} (``\texttt{youtu}'' \texttt{AND} ``\texttt{be}'') & all & 49,087,406 & 320,751 & 4,542,397 & 91.53\% \\
        \bottomrule
    \end{tabular}
\end{table*}

\clearpage

\section{A detailed comparison with Sampson et al. (2015) on the rate limit messages}
\label{sec:si_threads}

To understand the contradiction, we investigate a sampled dataset crawled on Sep 08, 2015 (dubbed \textsc{Sampled '15}).
This dataset contains tweets mentioning YouTube video URLs.
We believe that the design of rate limit messages in \textsc{Sampled '15} is the same with that measured in~\citeAP{sampson2015surpassing}.
Differing from the observations we make upon the 2019 \cb and \yt datasets, we notice 2 major differences for \textsc{Sampled '15}, which imply that the rate limit messages were implemented differently back in 2015.

\begin{figure}[!htbp]
    \centering
    \includegraphics[width=0.7\textwidth]{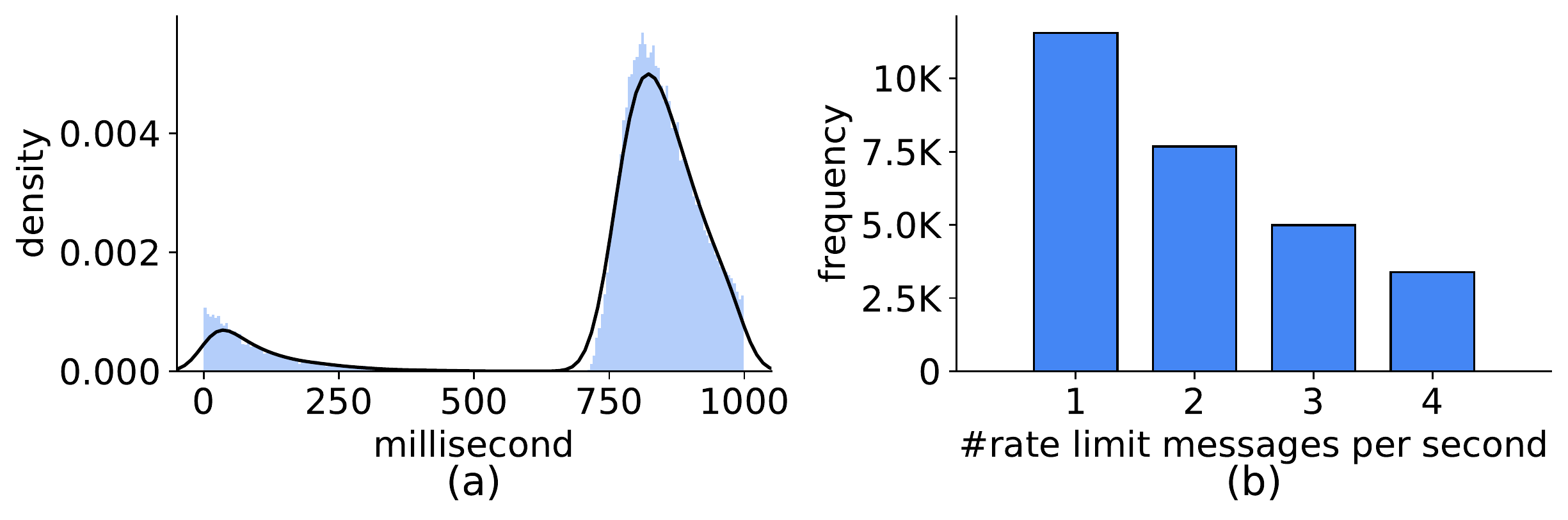}
    \caption{\textbf{(a)} The density distribution of milliseconds in the received 55,420 rate limit messages.
    \textbf{(b)} The number of rate limit messages in each second.}
	\label{fig:SI1}
\end{figure}

Firstly, \textsc{Sampled '15} receives up to 4 rate limit messages for each second.
\Cref{fig:SI1}(a) shows the milliseconds in the rate limit messages are not uniformly distributed -- 89\% rate limit messages are emitted between millisecond 700 to 1,000.
With a total of 55,420 rate limit messages, \Cref{fig:SI1}(b) shows that \{0,1,2,3,4\} rate limit messages can be received every second.
On the contrary, we obtain at most 1 rate limit message per second in the datasets crawled in 2019.

\begin{figure}[!htbp]
    \centering
    \includegraphics[width=0.7\textwidth]{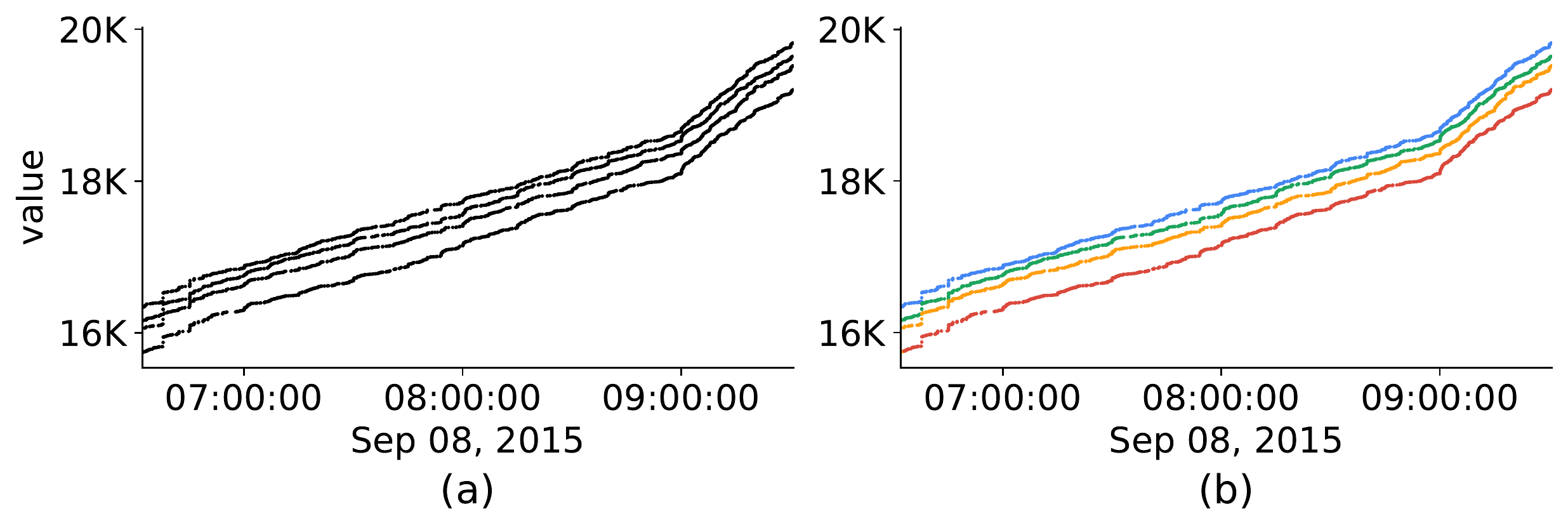}
    \caption{\textbf{(a)} Scatter plot of rate limit messages. x-axis: timestamp; y-axis: associated integer value.
    \textbf{(b)} Coloring rate limit messages into 4 monotonically increasing threads by using Algorithm~\ref{alg:mapping_ratemsg}.}
	\label{fig:SI2}
\end{figure}

Secondly, the integers in the rate limit messages are not increasing monotonically.
This contradicts Twitter's official documentation that ``\textit{Limit notices contain a total count of the number of undelivered Tweets since the connection was opened}~\citeAP{twitterrate1}''.
\Cref{fig:SI2}(a) shows the scatter plot of rate limit messages with the timestamp on the x-axis and the associated integer value on the y-axis.
The above observations prompt us to believe that the rate limit messages (and streaming clients) are split into 4 parallel threads rather than 1.
When the received messages are less than 4, one explanation could be that particular thread has delivered all tweets within it.
We propose Algorithm \ref{alg:mapping_ratemsg}, which maps rate limit messages to 4 monotonically increasing lists for estimating the number of undelivered tweets.
We color the mapping results in \Cref{fig:SI2}(b).
From the 4,500 rate limit messages received between 2015-09-08 06:30 UTC and 2015-09-08 09:30 UTC, we estimate that 85,720 tweets are missing.

\begin{algorithm*}[tbp]
  \SetKwInOut{Input}{input}\SetKwInOut{Output}{output}
  
  \Input{rate limit messages}
  \Output{multiple monotonically increasing lists \textbf{A} that consist of rate limit messages}
  \textit{initialize the returned list of list} \textbf{A} \textit{by the first element of rate limit messages}\;
  \While{not at end of messages}{
    read the next rate limit message $NextRate$\;
    read the last element from each of the existing lists \textbf{A} into $PrevRates$\;
    \uIf{$NextRate \leq min(PrevRates)$}{
      append a new list $[NextRate]$ to \textbf{A}\;
    }
    \uElse{
      $ 0\longleftarrow BestIndex$\;
      $ n \longleftarrow len(PrevRates) $\;
      \For{$k\longleftarrow 0$ \KwTo $n$}{
        \uIf{$ PrevRates[k] < NextRate < PrevRates[BestIndex] $}{
          $ k\longleftarrow BestIndex$\;
          }
      }
      update \textbf{A}$[BestIndex]$ by $NextRate$\;
    }
  }
  \caption{Mapping rate limit messages into multiple monotonically increasing lists.}
  \label{alg:mapping_ratemsg}
\end{algorithm*}

\begin{figure*}[!htbp]
    \centering
    \includegraphics[width=0.95\textwidth]{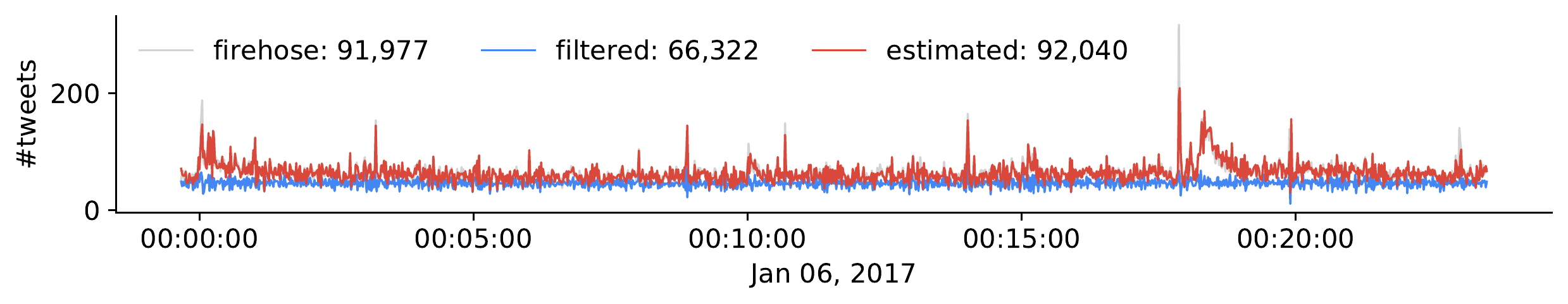}
	\caption{Comparing Twitter public filtered stream to the Firehose stream. MAPE is 0.68\% with estimated missing tweet volume from rate limit messages.}
	\label{fig:SI3}
\end{figure*}

To validate the 4-thread counters for rate limit messages, we obtain a complete data sample from a Twitter data reseller \textbf{discovertext.com}\footnote{\url{https://discovertext.com/}}.
This company provides access to the Firehose service at a cost.
We started two streaming clients simultaneously on 2017-01-06, one with the Twitter filtered streaming API, the other with the Firehose.
We track the temporal tweet volumes at three level: (1) public filtered stream (\textcolor{blue}{blue}); (2) filtered stream plus the estimated missing volume from rate limit messages (\textcolor{red}{red}); (3) the complete tweet stream from Firehose (grey).
The missing volume is estimated by using Algorithm \ref{alg:mapping_ratemsg}, which accounts the integers in rate limit messages onto 4 parallel counters.
As shown in~\Cref{fig:SI3}, the \textcolor{red}{red} line and grey line almost overlap each other (MAPE is 0.68\%).
Our experiments show that the 4-thread counters are accurate measures for the missing tweet volume in early 2017.e
Altogether, the collected tweets from public filtered stream plus the estimated missing tweets are close to both Firehose stream (as evident here) and tweets crawled from multiple subcrawlers (as evident in Section 3 of the main paper).
This eases our core assumption that multiple subcrawlers capture the underlying universe of Twitter Firehose stream.

If one uses the 1-thread counter to compute the missing volume, the total volume will be undercounted as the estimated missing volume now reduces to about 25\%.
This in turn makes the complete data stream appears to be much larger than the estimated total volume, which was what the researchers found in~\citeAP{sampson2015surpassing}.
Based on the above observations, we believe that the discrepancies between our work and \citeAP{sampson2015surpassing} are a result of different implementation choices of the rate limit messages in 2015 and 2019.

\clearpage

\section{Minutely and secondly sampling rates}
\label{sec:si_rates}

\begin{figure}[!htbp]
    \centering
    \includegraphics[width=0.7\textwidth]{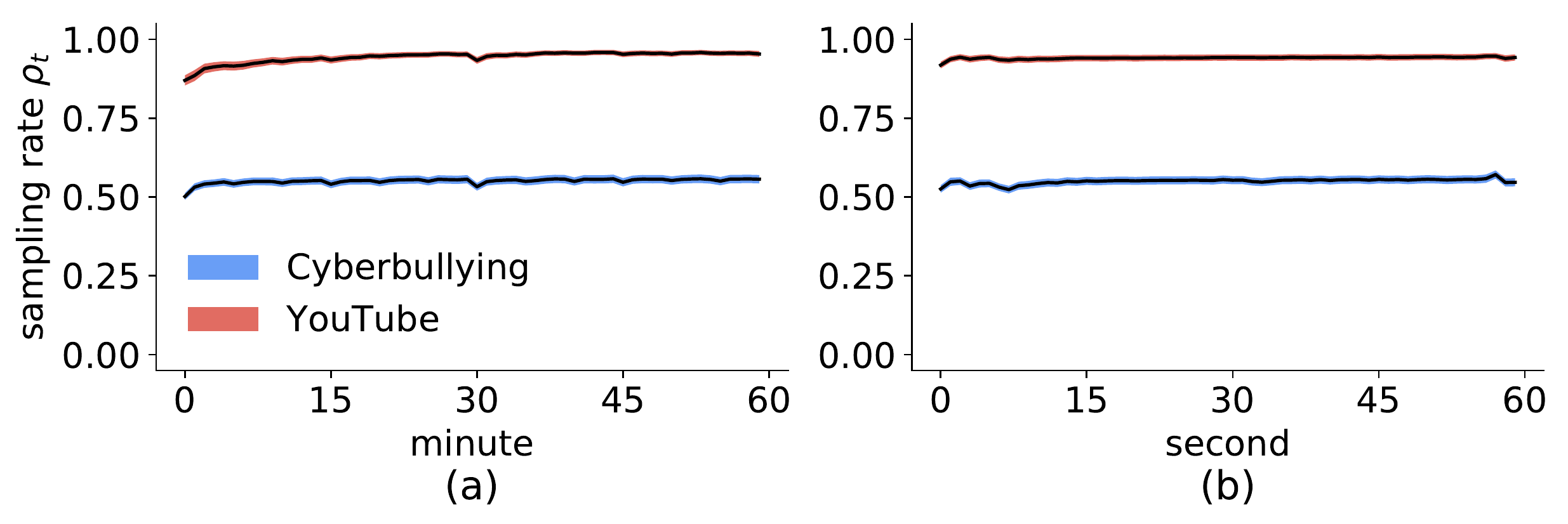}
	\caption{Minutely and secondly sampling rates.}
	\label{fig:SI4}
\end{figure}

The temporal variations are much less prominent at the minutely and secondly levels, as shown in \Cref{fig:SI4}.

\section{Universal rank percentiles change}
\label{sec:si_rank}

\begin{figure}[!htbp]
    \centering
    \includegraphics[width=0.3\textwidth]{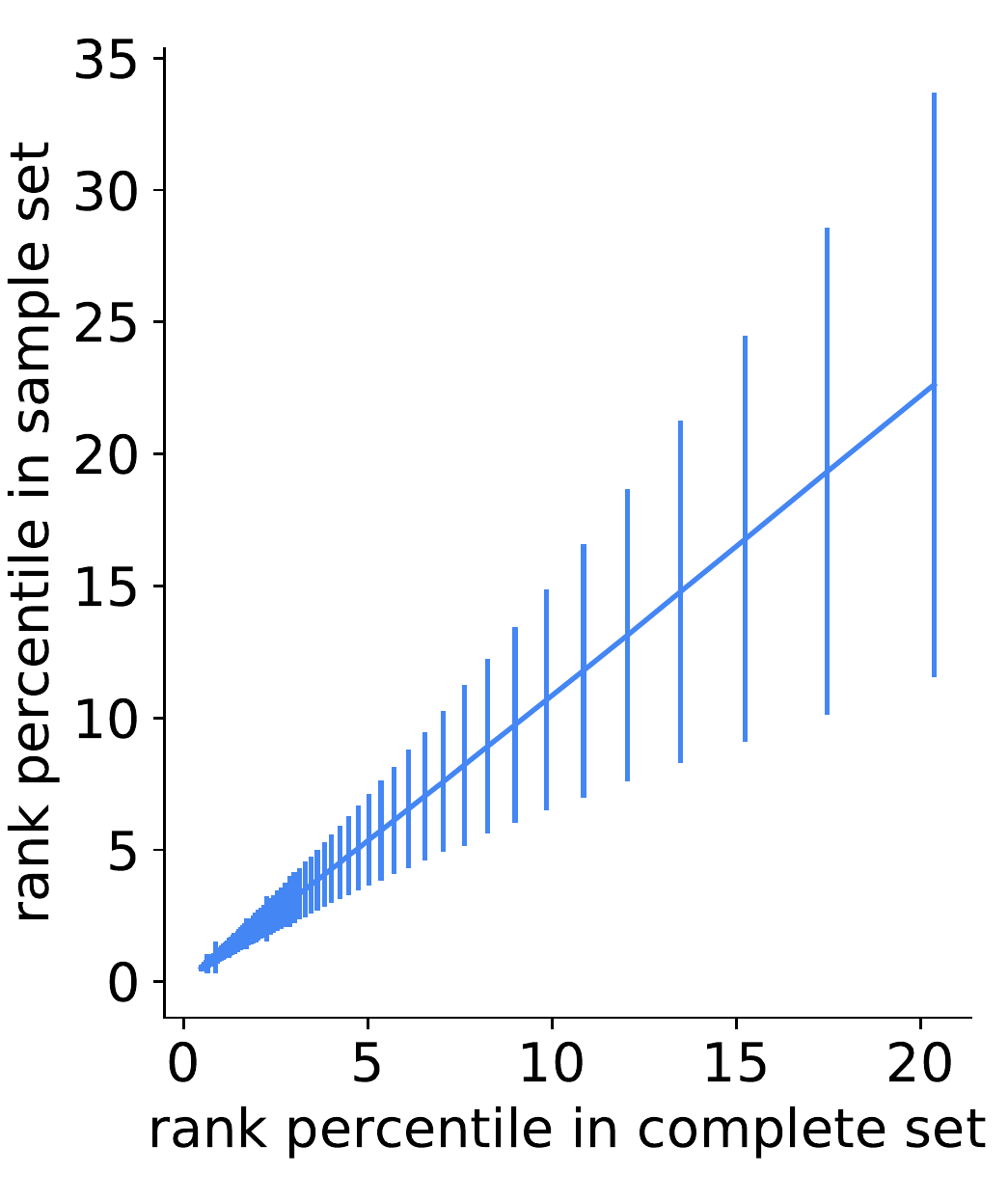}
	\caption{x-axis: true rank percentiles, y-axis: observed rank percentiles.}
	\label{fig:SI5}
\end{figure}

Concerning all users in our dataset (not just the most active 100 users), \Cref{fig:SI5} shows the true rank percentiles on the x-axis and the observed rank percentiles on the y-axis.
The smaller the value is, the higher rank the user posits among the cohort.
For an user that tweets 5 times in the complete set (displayed as the rightmost vertical line), their tweets appear on average 2.65 times in the sample set.
The true rank percentile is 20.37\%, and the observed rank percentile is 22.63\% $\pm$ 11.07\%.
For an user that tweets 40 times in the complete set, their tweets appear on average 21.09 times in the sample set.
The true rank percentile is 2.03\%, and the observed rank percentile is 2.14\% $\pm$ 0.58\%.
With a decreasing standard deviation, the percentiles of the higher ranked entities are more trustworthy.

\bibliographyAP{twitter-sampling-ref}
\bibliographystyleAP{aaai}

\end{document}